\documentclass{emulateapj}

\shorttitle{Turbulence in Dust Layer}
\shortauthors{Takeuchi et al.}
\slugcomment{Accepted by the Astrophysical Journal}

\begin{document}

\title{Induced Turbulence and the Density Structure of the Dust Layer in a Protoplanetary Disk}
\author{Taku Takeuchi\altaffilmark{1}, Takayuki Muto\altaffilmark{1}, Satoshi
Okuzumi\altaffilmark{2}, Naoki Ishitsu\altaffilmark{3}, and Shigeru Ida\altaffilmark{1}}

\begin{abstract}
We study the turbulence induced in the dust layer of a protoplanetary disk
based on the energetics of dust accretion due to gas drag. 
We estimate turbulence strength from the energy supplied by dust accretion, using the
radial drift velocity of the dust particles in a laminar disk. 
Our estimate of the turbulence strength agrees with previous analytical and
numerical research on the turbulence induced by Kelvin-Helmholtz and/or
streaming instabilities for particles whose stopping time is less than the
Keplerian time. For such small particles, the strongest turbulence is expected
to occur when the dust-to-gas ratio of the disk is $\sim C_{\rm
  eff}^{1/2} (h_g/r) \sim10^{-2}$, where $C_{\rm eff} \approx 0.2$
represents the energy supply efficiency to turbulence and $h_g/r\sim 5
\times 10^{-2}$ is the aspect ratio of the gas disk. The 
maximum viscosity parameter is $\alpha_{\mathrm{max}} \sim C_{\rm eff}
T_s (h_g/r)^2 \sim 10^{-4} T_s$, where $T_{s}(<1)$ is the non-dimensional
stopping time of the dust particles. Modification in the dust-to-gas
ratio from the standard value, $10^{-2}$, by any process, results in
weaker turbulence and a thinner dust layer, and consequently may
accelerate the growth process of the dust particles.
\end{abstract}


\altaffiltext{1}{Department of Earth and Planetary Sciences, Tokyo Institute of Technology, Meguro-ku, Tokyo, 152-8551, Japan; taku@geo.titech.ac.jp}
\altaffiltext{2}{Department of Physics, Nagoya University, Nagoya, Aichi 464-8602, Japan}
\altaffiltext{3}{National Astronomical Observatory, Mitaka, Tokyo 181-8588, Japan}
\setcounter{footnote}{3}

\section{INTRODUCTION}

The first step of planet formation in protoplanetary disks is the collisional
growth of (sub-)micron-sized dust particles (or aggregates), and consequent
sedimentation. Dust settling and the formation of a dust layer at the midplane of
the disk play an important role in the subsequent planetesimal formation
process. Enhancement of the particle density in the dust layer accelerates the
collisional growth rate. If the density enhancement is high enough,
planetesimals may form through the gravitational instability of the dust layer
(Goldreich \& Ward 1973; Sekiya 1983). However, the turbulent motion of the gas
hinders the dust layer from thinning. Dust particles are stirred by
turbulence and diffuse to high altitudes from the midplane. Thus, turbulence 
is an impediment to planetesimal formation. Turbulence excited by
magneto-rotational instability (MRI) is so strong that the dust layer cannot
become thin enough to induce planetesimal formation through gravitational
instability (Johansen \& Klahr 2005; Fromang \&\ Papaloizou 2006; Turner et
al. 2006; Carballido et al. 2006, 2011; Fromang \& Nelson 2009).  In addition, the
turbulence-induced collisional velocity can be high enough to destroy dust
aggregates (Carballido et al. 2008, 2010), though several mechanisms have been
proposed to overcome this difficulty (Johansen et al. 2007, 2011; Lyra et al.
2008, 2009). Even if the gas disk exhibits an initially laminar flow (such as is expected
in the dead zone where the ionization fraction of the gas is too low to couple
to the magnetic field, as mentioned in Gammie 1996 and Sano et al. 2000), dust 
settling itself induces turbulence. The dust particles tend to rotate around 
the central star faster than the gas because the gas experiences a pressure gradient 
force acting in the opposite direction of the gravity of the star. When the dust layer 
thins and the dust-to-gas ratio in the layer approaches unity, the velocity
difference from that of the upper gas layer induces Kelvin-Helmholtz
(KH) instability and excites turbulence in the dust layer (e.g.,
Goldreich \& Ward 1973; Sekiya \& Ishitsu 2000, 2001; Garaud \&\ Lin
2004). The velocity difference between the dust particles and the gas
inside the dust layer also induces streaming instability (Youdin \&
Goodman 2005; Youdin \& Johansen 2007; Johansen \& Youdin 2007; see
Chiang \& Youdin 2010 for review on the various instability in the dust
layer). Laminar gas disks are considered excellent sites for planetesimal
formation.  Thus, it is important to clarify how turbulence induced in
the dust layer diffuses the dust particles, versus dust
sedimentation. Several analytical and numerical studies have focused on
this problem (Cuzzi et al. 1993; Champney et al. 1995; Sekiya 1998;
Dobrovolskis et al. 1999; Johansen et al. 2006; Michikoshi \&\ Inutsuka
2006; Weidenschilling 2006, 2010; Bai \&\ Stone 2010a, 2010b).

Sekiya (1998, hereafter S98) analytically solved for the structure of the dust
layer under turbulence induced by KH instability. To derive the density
profile of the dust layer analytically, the author adopted several assumptions.  First, the dust particles were assumed to be small enough and were coupled so tightly to the gas that
the dust and the gas could be treated as a single fluid.  Second, the density
structure was adjusted to keep the dust layer marginally unstable
to KH instability. Third, the effects of the Coriolis force and of the Keplerian
shear were neglected. These assumptions allowed the author to determine
analytically the density structure of the dust layer and to discuss how far
from the midplane the dust particles diffuse due to turbulence. The
analysis in S98 provided a useful guide for subsequent numerical studies
that investigated turbulence of the dust layer with more realistic assumptions.
In reality, however, dust particles that are not tightly coupled to gas
play an important role in inducing turbulence. Also, other kinds of instability,
such as streaming instability, may occur before KH instability sets in (Bai \&
Stone 2010a). Thus, in addition to performing numerical simulations, it is
desirable to include an analytic discussion on the structure of the dust layer using
a more general set of assumptions than adopted in S98.

In this paper, we revisit the analysis presented in S98 from a different point of view.
In S98, the density structure of the dust layer was determined by the condition
that the dust layer was marginally unstable to KH instability; i.e., the
Richardson number, $\mathrm{Ri}$, had a constant critical value. In the
stability condition on $\mathrm{Ri}$, the free energy due to the shear
velocity is compared with the energy needed to lift the material against the
vertical gravity (Chandrasekhar 1961). In the case of a dust layer, the free
energy of the azimuthal velocity is compared with the gravitational potential in
the vertical direction. We also discuss the energetics of the dust layer to
determine the density structure, but in this paper we focus on the
gravitational potential in the radial direction. Sustaining steady turbulence
requires an energy supply. The source of the free energy for KH instability or
streaming instability is the velocity difference between the dust and the gas.
Because this free energy is consumed in the process of exciting turbulence, a velocity difference must be continually induced to provide steady turbulence. This velocity difference originates from the force balance in the radial direction
between the stellar gravity, the centrifugal force, and the pressure gradient.
In other words, the velocity difference is induced because the gas resides in a slightly shallower effective potential (including both the
stellar gravity and the gas pressure), than the effective potential for the dust. The dust
loses its angular momentum due to gas drag and drifts towards the star, while the gas gains 
angular momentum and drifts outward. The total angular momentum remains
constant. Because the effective potential for the dust is deeper, this process releases 
gravitational energy that can be a source of the free energy for turbulence. (The relationship between the angular momentum variation, $\Delta L$, and the energy
variation, $\Delta E$, for circular orbiting material is $\Delta E=\Omega\Delta
L$, where the angular velocity, $\Omega$, is different for the gas and for the
dust.) If the gas disk is initially in a state of laminar flow, i.e., if gas
drag is the only process that exchanges angular momentum, then dust accretion is the source of the free energy used to induce turbulence. The accretion
rate of the dust due to gas drag has been calculated in the literature
(Nakagawa et al. 1986, hereafter NSH86; Weidenschilling 2003; Youdin \& Chiang
2004). Using this dust accretion rate, and assuming that a certain
fraction of the accretion energy is transferred to turbulence, it is
straightforward to calculate the strength of the turbulence and to determine the
density structure of the dust layer. In this paper, we show that an analysis of the
energetics based on the gravitational potential in the radial direction results in a dust layer structure that is qualitatively equivalent to the dust layer structure developed in
the analysis presented in S98. Most of the important properties of the S98 model
are reproduced. Our analysis avoids some of the assumptions that were adopted in S98.  
Specifically, we do not assume tight coupling between the gas and the dust, nor a
marginally unstable structure for KH instability. (The effects of the Coriolis
force and of Keplerian shear, which are still neglected in this paper, are
discussed by Ishitsu \& Sekiya 2002, 2003; G\'omez \& Ostriker 2005; Chiang
2008; Barranco 2009; Lee et al. 2010.) Thus, the results of this paper may be applied to more general situations, including situations in which the
dust is weakly coupled to the gas, or in which streaming instability acts as a source
of turbulence.

In \S \ref{sec:assump}, we describe our model assumptions. In
\S \ref{sec:eneinp}, the energy release rate of the accreting dust is
calculated. In \S \ref{sec:turbstr}, the strength of the turbulence is derived and
its asymptotic forms in the limits of a small dust-to-gas ratio and a
tight dust-gas coupling are discussed. In \S \ref{sec:comppre}, the results of
our model are compared with the S98 model and with previous numerical
simulations of KH instability and streaming instability. In
\S \ref{sec:dustvelocity}, we discuss how dust layer formation reduces
the radial drift velocity of the dust and the relative velocity of the dust
particles. In \S \ref{sec:ene-into-tur}, we check the applicability of our model to
turbulence in the dust layer for the KH instability and streaming instability cases.

\section{DISK MODEL AND ASSUMPTIONS}
\label{sec:assump}

\subsection{Brief Model Description}

We consider a dust layer that forms after dust particles have settled to the
midplane of a gas disk around a star. Before the dust settles, the gas disk is
assumed to be subject to laminar flow. This means that we consider the dead zone where MRI is
inactive because of the low degree of ionization. Turbulent gas motion may be present even in the dead zone because sound waves propagate from the
active layers at high altitudes and disturb the gas in the dead zone (Fleming \&
Stone 2003; Suzuki et al. 2010; Okuzumi \& Hirose 2011). We assume that the
dead zone is wide enough that the active layer cannot induce strong turbulence
at the midplane.

Even in an initially laminar disk, dust sedimentation and the formation of the
midplane dust layer cannot proceed in a  perfectly undisturbed fashion. 
When the midplane dust-to-gas
ratio reaches a critical value, the velocity differences between the dust
layer and the upper gas layer, or between the individual dust particles and the
surrounding gas, will start to induce hydrodynamical instabilities, such as KH
instability or streaming instability (see Chiang \& Youdin 2010 for review).
Consequently, turbulent diffusion of the dust particles terminates further
dust settling. It is expected that a steady dust layer forms, in which 
particle settling and turbulent diffusion balance each other. In such a
state, vertical settling is no longer an energy source for turbulence,
but the radial accretion of the dust still proceeds. The turbulence is
maintained by the energy liberated from dust accretion to the star.

We calculate the dust accretion rate and consequent energy release, 
assuming that a steady dust layer has formed. The strength of the
turbulence is then estimated. The balance between the turbulent diffusion
and the dust settling determines the dust layer thickness. Thus, the structure of
the dust layer, the dust accretion rate, and the turbulence strength must be
determined self-consistently.

For simplicity, we focus on a narrow axisymmetric ring region at a certain
radius $r$ from the central star. We consider the dust layer structure only in
the vertical direction, neglecting any radial variation in the properties of
the gas disk and the dust layer. All the dust particles are assumed to have
a uniform size, i.e., the size distribution of the dust particles is neglected
in this paper.

\subsection{Vertical Structure of the Dust Layer}

If the effect of gas drag on the dust were negligible, the dust particles
 would orbit at the Keplerian velocity 
$\Omega_{\mathrm{K}}=\sqrt{(GM)/r^{3}}$, where $G$ is the 
gravitational constant and $M$ is the mass of the
central star. However, the orbital velocity of the gas is slightly less
than the Keplerian velocity because the direction of the gas pressure
gradient is usually outward from the star.  This cancels part of the 
contribution to the orbital velocity from the gravity of the star. Given a negligible influence of 
the drag force from the dust, the angular velocity of the gas would be
\begin{equation}
\Omega_{g}=\sqrt{\frac{GM}{r^{3}}(1-2\eta)}\approx\Omega_{\mathrm{K}}(1-\eta).
\label{eq:omega_g}%
\end{equation}
Here, $\eta$ represents the deviation of the gas orbital velocity from the
Keplerian value. It is half of the ratio between the gas pressure gradient
force and the gravity of the central star,
\begin{equation}
\eta=-\frac{1}{{2\rho_{g}r\Omega_{\mathrm{K}}^{2}}}\frac{{\partial P}}{{\partial r}}\ ,
\end{equation}
where $\rho_{g}$ and $P$ are the gas density and pressure, respectively.

We consider a thin dust layer in which most of the dust particles have settled
on the midplane of the gas disk. The gas disk is isothermal in the vertical
direction with a scale height of $h_{g}=c_{s}/\Omega_{\mathrm{K}}$, where
$c_{s}$ is the gas sound speed. Because the thickness of the dust layer is
much smaller than the scale height of the gas disk, the gas density $\rho_{g}$
in the dust layer is treated as a constant. Its value is given by the column
density of the gas disk $\Sigma_{g}$ as
\begin{equation}
\rho_{g}=\frac{\Sigma_{g}}{\sqrt{2 \pi} h_{g}}\ . 
\label{eq:rhog}%
\end{equation}
The dust density profile in the vertical direction, $\rho_{d}(z)$, is
determined by the balance between the dust settling and turbulent diffusion
(Youdin \& Lithwick 2007, hereafter YL07)
\footnote{Note that the diffusion coefficient
given by YL07 is derived assuming that the dust particles can be treated as
passive particles. Problems may arise if this formula is used in a case 
where the dust layer is so dense that its dust-to-gas ratio is larger than unity. Although a
comparison of our model with the simulation by Johansen et al. (2006) shows 
good agreement even for a dust-to-gas ratio as large as $40$ (see Fig.
\ref{fig:fdmid-joh}), modeling the effect of the dust inertia will be a 
subject for future investigation using a refinement of the present
model.}.
The dust particles have a
stopping time, $\tau_{\mathrm{stop}}=T_{s}\Omega_{\mathrm{K}}^{-1}$, in which
the velocity difference from the background gas flow becomes $1/e$ times due
to gas drag, where $T_{s}$ is the non-dimensional stopping time. In a
turbulent gas disk with a turbulent diffusion coefficient $D_g$,
the density profile of the dust layer composed of single-sized particles is
Gaussian, $\rho_{d}=\rho_{d,0}\exp(-\tilde{z}^{2})$, where $\tilde{z}%
=z/(\sqrt{2}h_{d})$ is the non-dimensional vertical coordinate normalized by
the dust scale height $h_d \approx [D_g /(\Omega_{\rm K} T_s)]^{1/2}$.
Note that the turbulent diffusion coefficient in the ``z-direction'',
$D_g$, may have a significantly different value from the usual turbulent
viscosity coefficient $\nu_{\rm acc}$ in accretion disks, where
$\nu_{\rm acc}$ comes from the ``$r \theta$-component'' of the Reynolds
stress (e.g., Lesur \& Ogilvie 2010). Nevertheless, we follow the conventional
``$\alpha$-prescription'', and express the diffusion coefficient as
$D_g=\alpha c_g h_g$ for simplicity. This prescription and equation (24)
of YL07 give
\begin{equation}
h_{d}=\sqrt{\frac{\alpha}{{T_{s}}}}\sqrt{\frac{{1+T_{s}}}{{1+2T_{s}}}}h_{g}\ .
\label{eq:hd}%
\end{equation}
The dust density profile is written as
\begin{equation}
\rho_{d}=f_{\mathrm{mid}}\rho_{g}\exp(-\tilde{z}^{2})\ ,
\end{equation}
where the midplane dust-to-gas ratio $f_{\mathrm{mid}}$ is increased by a
factor $h_{g}/h_{d}$ from the total dust-to-gas ratio or the
``metallicity'' of the disk $Z_{\mathrm{disk}}=\Sigma_d / \Sigma_g$,
\begin{equation}
f_{\mathrm{mid}}=Z_{\mathrm{disk}}\frac{{h_{g}}}{{h_{d}}}=Z_{\mathrm{disk}}\sqrt{\frac{{T_{s}}}{\alpha}}\sqrt{\frac{{1+2T_{s}}}{{1+T_{s}}}}\ .
\label{eq:fdmid}%
\end{equation}
We define $\beta(\tilde{z})$ as the ratio of the total (dust and gas)
density to the gas density,
\begin{equation}
\beta=\frac{\rho_{g}+\rho_{d}}{\rho_{g}}={1+f_{\mathrm{mid}}\exp(-\tilde
{z}^{2})}\ . 
\label{eq:beta}%
\end{equation}

\subsection{Typical Values of the Model Parameters}

In our model, the turbulent parameter $\alpha$ and the structure of the dust
layer (i.e., the thickness, $h_{d}$, and the midplane dust-to-gas ratio,
$f_{\mathrm{mid}}$) are determined for a given parameter set (the stopping
time of the dust particles $T_{s}$, the disk metallicity
$Z_{\mathrm{disk}}$, and the parameters for the gas disk). The
non-dimensional variables of the result ($\alpha$, $h_{d}/r$, and
$f_{\mathrm{mid}}$) depend on the gas disk parameters only through
\begin{equation}
\tilde{\eta}=\eta^{2}\left(  \frac{r}{h_{g}}\right)  ^{2}~, 
\label{eq:etawave}%
\end{equation}
where $\tilde{\eta}\sim\eta\sim(h_{g}/r)^{2}\sim10^{-3}-10^{-2}$. The
numerical calculations in the subsequent sections mainly use a value
$\tilde{\eta}=(0.05/\gamma)^{2}=9 \times 10^{-4}$, where $\gamma=5/3$,
in order to compare our model with the simulation presented by 
Johansen et al. (2006, hereafter JHK06). 
In \S\ref{sec:BS10}, $\tilde{\eta}=0.05^{2}$ is used for comparison
with the results by Bai \& Stone (2010), and in \S\ref{sec:dustvelocity},
$\tilde{\eta}=2.92 \times 10^{-3}$ is used to calculate the radial
drift velocity  of the dust in the gas disk model by Hayashi (1981). The
$\tilde{\eta}$ dependence of the results will be discussed in
\S\ref{sec:disketa}.

\section{TURBULENCE ENERGY SUPPLY}
\label{sec:eneinp}

The energy supply for turbulence comes from liberation of the gravitational
energy of the dust. As the dust falls towards the central star, the dust
particles penetrate more deeply into the potential well of the star. Although the gas drifts
outward to conserve the total angular momentum, the difference in the
effective gravitational potential between the dust ($-GM/r$) and the gas
($-GM(1-2\eta)/r$), including the work done by the pressure gradient,
causes a net energy liberation. Part of the liberated energy is converted
directly into thermal energy, and part is used for supplying energy to
the turbulence (see discussion in \S \ref{sec:ene-into-tur}).

Dust particle accretion occurs either individually or
collectively. Individual dust particles suffer gas drag given a
velocity difference between the particle and the gas. Usually the orbital
velocity of the gas is slower than that of the dust particle, and the gas drag
force decelerates the particle orbital motion. Thus, the particle loses its
angular momentum and drifts inward. In addition to this individual drift,
a collective drag works on the whole dust layer. The dust layer
at the midplane orbits faster than the gas layer at higher altitudes because in the
dust layer the increased inertia of the enriched dust particles weakens the
effect of the gas pressure. The dust layer rotates with a velocity close to
the Keplerian value. The upper gas layer, devoid of the inertia of the dust,
orbits more slowly than the midplane dust layer. If the gas exhibits turbulent
viscosity, then the slower orbiting gas layer exerts a drag force on the dust
layer, and consequently the dust layer loses its angular momentum and accretes
inward. We consider the energy release rate of the dust caused by 
individual and collective drag.

\subsection{Dust Accretion Caused by Individual Drag}
\label{sec:inddrag}

We calculate the gravitational energy released when dust
particles accrete towards the star due to drag on individual particles. In this
calculation, we assume that the gas disk is in a laminar flow state, and that
dust particles drift inward steadily. Any random motion of particles caused by
gas turbulence is neglected. We expect that in turbulent gas
disks, the average motion of the dust particles can be estimated from the motion in
the laminar disk (Bai \& Stone 2010). Because the individual drag is more
effective than the collective drag when the midplane dust-to-gas ratio is less
than unity (see Fig. \ref{fig:Sigma} below), we take the $f_{\mathrm{mid}}%
\ll1$ limit in the following calculation. The calculation for a general $f_{\mathrm{mid}}$ is described in Appendix \ref{sec:app-ind}.

In a laminar disk, the particle drift velocity $v_{d,r}$ is given by equation
(2.11) of NSH86. For $\rho_{d}\ll\rho_{g}$, the particle radial velocity
is (see also eq. [23] of Takeuchi \& Lin (2002); note the factor of 2
difference in the definition of $\eta$)%
\begin{equation}
v_{d,r}=-\frac{2{T_{s}}}{{T_{s}^{2}+1}}\eta v_{\mathrm{K}}\ . 
\label{eq:vdr}%
\end{equation}
The gas drifts in the opposite direction with a velocity $v_{g,r}$. From
angular momentum conservation, the gas drift velocity is
\begin{equation}
v_{g,r}=-\frac{{\rho_{d}}}{{\rho_{g}}}v_{d,r}\ . 
\label{eq:vgr}%
\end{equation}
The effective gravities (including the pressure gradient force) acting on the
dust and on the gas are $g_{d}=-GM/r^{2}=-r\Omega_{\mathrm{K}}^{2}$ and
$g_{g}=-GM(1-2\eta)/r^{2}=-r\Omega_{g}^{2}$, respectively. While the total
angular momentum is conserved, the total energy is not conserved because of the
difference in the effective gravities acting on the gas and on the dust,
$\rho_{d}{g_{d}v_{d,r}+\rho}_{g}g_{g}v_{g,r}{\neq0}$, providing a source of
energy for turbulence.

Note that, in the above calculation, $v_{d,r}$ and $v_{g,r}$ are the
\textquotedblleft terminal velocities\textquotedblright. This means that the
gravitational accelerations of the dust and gas are balanced by the drag
forces, and liberated energy translates directly into thermal energy. Thus,
no energy would be contributed to turbulence. In reality, however, if the gas
disk is turbulent, a steady terminal velocity is not expected, and acceleration phases of the dust must occur, as
discussed in \S \ref{sec:ene-into-tur}. In an acceleration phase, the work
done by gravity first provides kinetic energy, which can be translated
into turbulent energy. We estimate this energy input to turbulence. In the
following calculation, the energy liberated from the accreting dust is
estimated using the terminal velocity described above for simplicity. Only a
certain fraction of the liberated energy goes into turbulence. Thus, the
energy input to turbulence is a factor $C_{\mathrm{eff}}(<1)$ times the
following estimation. The factor $C_{\mathrm{eff}}$ will be
determined in \S \ref{sec:comp-sim} to be $\approx0.2$ by comparing our
model with the numerical simulation of turbulence in the dust layer
developed by JHK06.

Given that some portion of the accretion energy of the dust is consumed by the outward motion of the gas, the liberated gravitational energy per unit surface area of the disk is,%
\begin{equation}
\frac{\partial E_{\mathrm{drag}}}{\partial t}=\frac{1}{2}\int_{-\infty
}^{\infty}(\rho_{d}{g_{d}v_{d,r}+}\rho_{g}{g_{g}v_{g,r})dz}~,
\end{equation}
where the factor of $1/2$ comes from the fact that half of the work done by
gravity is used for acceleration (and deceleration) of the azimuthal velocity
of the dust (and of the gas) as their semi-major axes change. Using equation
(\ref{eq:vdr}) and (\ref{eq:vgr}), the energy liberation rate reduces to
\begin{equation}
\frac{\partial E_{\mathrm{drag}}}{\partial t}=2\eta^{2}v_{\mathrm{K}}%
^{2}\Omega_{\mathrm{K}}T_{s}\Sigma_{d,\mathrm{drag}}\ , 
\label{eq:ene_drag}%
\end{equation}
where the effective ``surface density'' $\Sigma_{d,\mathrm{drag}}$ of the dust
is%
\begin{equation}
\Sigma_{d,\mathrm{drag}}=\frac{1}{{T_{s}^{2}+1}}\Sigma_{d}~.
\label{eq:sig_drag2}%
\end{equation}
In the above calculation, we assume $f_{\mathrm{mid}}\ll1$. The calculation of
the energy liberation rate for general $f_{\mathrm{mid}}$ is described in
Appendix \ref{sec:app-ind}. In the numerical calculations in the subsequent
sections, we use equation (\ref{eq:sig_drag_ap}) for the effective surface
density
\footnote{Equation (\ref{eq:sig_drag}) does not coincide with
equation (\ref{eq:sig_drag2}) in the limit of $f_{\mathrm{mid}}\ll1$. This
discrepancy comes from the fact that the calculation of equation
(\ref{eq:sig_drag}) includes correction terms of
the order of $\eta^{2}$ in the drift velocity, while equation
(\ref{eq:sig_drag2}) considers only the terms of $\eta$. The difference
between equations (\ref{eq:sig_drag2}) and (\ref{eq:sig_drag}) is at most
factor 2 (for $T_{s}\gg1)$, and is not significant.},
\begin{equation}
\Sigma_{d,\mathrm{drag}}=\frac{{\Sigma_{d}}}{\sqrt{\pi}}\int_{-\infty}%
^{\infty}{\frac{{\exp(-\tilde{z}^{2})}}{{T_{s}^{2}+\beta^{2}}}\left[
1-\frac{T_{s}^{2}}{2(T_{s}^{2}+\beta^{2})}\right]  d\tilde{z}}\ .
\label{eq:sig_drag}%
\end{equation}
If the dust particles are tightly coupled to the gas ($T_{s}\ll1$) and
dust sedimentation is weak ($f_{\mathrm{mid}}\ll1$), then
$\Sigma_{d,\mathrm{drag}}$ is simply equal to the dust surface density
$\Sigma_{d}$.

\subsection{Dust Accretion Caused by Collective Drag}
\label{sec:coldrag}

Next, we consider dust accretion due to collective drag acting on the entire
dust layer. Because the faster-orbiting dust particles drag the gas in the
dust layer, the orbital velocity of the gas is largest at the midplane and
decreases with altitude. If the gas in the dust layer is turbulent, the
variation in orbital velocity $v_{g,\theta}$ with altitude $z$ induces
Reynolds stress $P_{\theta z}$. This causes a transfer of angular momentum from the dust
layer to the upper gas layer, resulting in accretion of the dust layer.

To calculate the energy liberation rate, we make a few key
assumptions. First, we consider only the $\theta z$-component of the Reynolds
stress, $P_{\theta z}$, neglecting the other components $P_{r\theta}$ and
$P_{zr}$. Ignoring $P_{r\theta}$ means that turbulence in the dust layer does
not transfer the angular momentum efficiently in the radial direction. This is
expected for turbulence induced by hydrodynamical instabilities such as
convective instability (e.g., Stone \& Balbus 1996; Lesur \& Ogilvie 2010).
The Reynolds stress $P_{zr}$ is also neglected for simplicity. Brauer et al.
(2007) pointed out that $P_{zr}$ changes the velocity profiles of the dust and
gas from those derived by NSH86 by a factor $\sim3$, and thus $P_{zr}$
cannot be neglected in a rigorous discussion.
Obtaining the exact velocity profiles including $P_{zr}$ requires
numerical calculations. In this paper, we use the velocity profiles calculated
analytically by NSH86 for simplicity. This induces an error of a factor
$\sim3$ in our estimation of turbulence strength. Thus, our estimate is
limited to an order-of-magnitude argument.

The second assumption is that the turbulent layer has a thickness
comparable to that of the dust layer. If the turbulent layer were much thicker
than the dust layer and most of the volume of the turbulent layer were free of
the dust, then its structure would be controlled by the gas, 
unaffected by the properties of the dust and the structure of the dust layer.
The dust layer would behave just like a boundary wall at the bottom of
the turbulent layer. Turbulence in such a thick boundary gas layer has
been discussed using the analogy of the Ekman layer (e.g., Cuzzi et al. 1993). 
If the turbulent layer were dominated by the
dust layer, then the structure of the dust layer would control the
turbulence strength. We focus on such a dusty turbulent layer. Following
Youdin \& Chiang (2004), we consider the conditions needed for the
turbulent and dust layers to be of similar thickness. The thickness of
the turbulent layer, from dimensional analysis, is the Ekman length,
\begin{equation}
h_{E}\sim\sqrt{\frac{\nu}{\Omega_{\mathrm{K}}}}~, 
\label{eq:he}%
\end{equation}
where $\nu$ is the turbulent viscosity, provided that the viscosity and the
Coriolis force determine the layer structure. The thickness
of the dust layer (eq. [\ref{eq:hd}]), which is determined by the balance
between sedimentation and diffusion of the dust particles, is
\begin{equation}
h_{d}\sim\sqrt{\frac{\nu}{T_{s}\Omega_{\mathrm{K}}}}~, 
\label{eq:hd2}%
\end{equation}
where $\nu \sim D_g$ is used.
For small particles ($T_{s}\la1$), $h_{d}$ is larger than $h_{E}$, meaning
that such small particles move further out of the turbulent layer and
modify the structure of the turbulent layer. It is expected that the thickness
of the turbulent layer is not given by $h_{E}$, but by $h_{d}$ (see also S98;
Goodman \& Pindor 2000). For large particles ($T_{s}\gg1$), the turbulent
layer is much thicker than the dust layer, and its thickness is expected to be
$h_{E}$. In the following discussion, we consider a dusty turbulent
layer with a thickness similar to $h_{d}$. Thus, the analysis in this
paper is probably not appropriate for large particles ($T_{s}\gg1$).

Under the above assumptions, the energy liberation rate is estimated. Because
the collective drag is effective only if the midplane dust-to-gas ratio is
larger than unity (see Fig. \ref{fig:Sigma} below), we consider the case in
which a dense dust layer has formed ($f_{\mathrm{mid}}\ga1$) and use the
plate drag approximation (Goldreich \& Ward 1973; Goodman \& Pindor 2000;
Weidenschilling 2003). The calculation for general $f_{\mathrm{mid}}$ is
described in Appendix \ref{sec:app-col}. The Reynolds stress $P_{\theta z}$
near the boundary between the dust layer (or the turbulent layer) and the gas
layer is estimated as
\begin{equation}
P_{\theta z}=\rho_{g}\nu\frac{{\partial v_{g,\theta}}}{{\partial z}}\sim
-\rho_{g}\nu\frac{\eta v_{\mathrm{K}}}{h_{d}}~. 
\label{eq:Ptz}%
\end{equation}
This stress extracts angular momentum from the dust layer and transfers it to
the gas layer. The unit surface of the dust layer loses angular momentum $\partial
L_{d}/\partial t=rP_{\theta z}$, and the corresponding energy change is $\partial
E_{d}/\partial t=\Omega_{\mathrm{K}}\partial L_{d}/\partial t$. The gas layer
gains the same amount of angular momentum $\partial L_{g}/\partial
t=-rP_{\theta z}$, but the energy change is different from that of the
dust layer because of the work done by the pressure gradient: $\partial
E_{g}/\partial t=\Omega_{g}\partial L_{g}/\partial
t=(1-\eta)\Omega_{\mathrm{K}}\partial
L_{g}/\partial t$. In total, the energy liberation rate (the minus sign is
added), using equations (\ref{eq:rhog}),(\ref{eq:hd}),(\ref{eq:fdmid}), and
(\ref{eq:Ptz}), is
\begin{equation}
\frac{\partial E_{\mathrm{vis}}}{\partial t}=-(\Omega_{\mathrm{K}}-\Omega
_{g})rP_{\theta z}=2\eta^{2}v_{\mathrm{K}}{}^{2}\Omega_{\mathrm{K}}T_{s}%
\Sigma_{d,\mathrm{vis}}~, 
\label{eq:ene_vis}%
\end{equation}
where
\begin{equation}
\Sigma_{d,\mathrm{vis}}=\frac{1}{2\sqrt{2\pi}}\frac{1+2T_{s}}{1+T_{s}}\frac
{1}{f_{\mathrm{mid}}}\Sigma_{d}~. 
\label{eq:sig_vis-approx}%
\end{equation}
In the above calculation, the viscosity coefficient is modeled as
$\nu=\alpha c_g h_g$. The same $\alpha$ is used for both $D_g$ and
$\nu$ for simplicity. 
The effective surface density $\Sigma_{d,\mathrm{vis}}$ is inversely
proportional to the midplane dust-to-gas ratio $f_{\mathrm{mid}}$ and only
weakly depends on $T_{s}$. Note that the above relationship is derived for
$f_{\mathrm{mid}}\ga1$. The effective surface density for general
$f_{\mathrm{mid}}$ is derived in Appendix \ref{sec:app-col} and is given by
\begin{eqnarray}
\Sigma_{d,\mathrm{vis}} &=&
\frac{C_{\mathrm{str}}}{\sqrt{{\pi}}}f_{\mathrm{mid}}\Sigma_{d}\frac{{1+2T}_{s}}{1+{T_{s}}}
\nonumber \\
& \times & \int_{-\infty}^{\infty}{\frac
{{\beta^{2}+2\beta T_{s}^{2}-T_{s}^{2}}}{{(T_{s}^{2}+\beta^{2})^{2}}}%
\frac{{\tilde{z}^{2}\exp(-2\tilde{z}^{2})}}{{[1+C_{\mathrm{str}}%
f_{\mathrm{mid}}\exp(-\tilde{z}^{2})]}}d\tilde{z}}\ , \nonumber \\
\label{eq:sig_vis}%
\end{eqnarray}
where
\begin{equation}
{C_{\mathrm{str}}=}\frac{1}{T_{s}^{2}+1}~.
\end{equation}
Expression (\ref{eq:sig_vis}) is complicated, but for large $f_{\mathrm{mid}}$, its dependence on $T_{s}$ and $f_{\mathrm{mid}}$ is similar to that of
the simpler equation (\ref{eq:sig_vis-approx}); $\Sigma_{d,\mathrm{vis}}\propto
f_{\mathrm{mid}}^{-0.9}$ and depends only weakly on $T_{s}$.

\subsection{Energy Dissipation in Turbulence}

The turbulent energy of the largest eddies transfers to smaller eddies,
and finally dissipates to thermal energy via decay of the smallest eddies due
to molecular viscosity. The size and velocity of the largest eddies are
assumed to be $l_{g,\mathrm{eddy}}=\alpha^{1/2}h_{g}$ and $u_{g,\mathrm{eddy}%
}=\alpha^{1/2}c_{s}$. This assumption means that the turnover time of the
largest eddies is the Keplerian time ($\tau_{g,\mathrm{eddy}}%
=l_{g,\mathrm{eddy}}/u_{g,\mathrm{eddy}}=\Omega_{\mathrm{K}}^{-1}$; Cuzzi et
al. 2001). Using the Kolmogorov scaling law, the energy dissipation rate per
unit volume and unit time is
\begin{equation}
\frac{\partial\varepsilon_{\mathrm{turb}}}{\partial t}=({\rho_{g}%
+C_{\mathrm{ene}}\rho_{d})}\frac{{u_{g,\mathrm{eddy}}^{2}}}{\tau
{_{g,\mathrm{eddy}}}}=({\rho_{g}+C_{\mathrm{ene}}\rho_{d})}\alpha h_{g}%
^{2}\Omega_{\mathrm{K}}^{3}\ ,
\end{equation}
where the factor $C_{\mathrm{ene}}$ represents the fact that dust
particles that are coupled only weakly to the gas do not contribute to
the turbulent energy, and is given by (see Appendix \ref{sec:app-ene-dis})
\begin{equation}
C_{\mathrm{ene}}=\left\{
\begin{array}
[c]{ccc}%
\frac{1}{T_{s}T_{e}+1} & \mathrm{for} & T_{s}\,\leq T_{e}\\
\frac{1}{T_{s}^{2}(T_{e}^{-2}+1)} & \mathrm{for} & T_{s}\,>T_{e}%
\end{array}
\right.  \ ,
\end{equation}
where $T_{e}=\tau_{g,\mathrm{eddy}}\Omega_{\mathrm{K}}$ is the non-dimensional
turnover time of the largest eddies. For most of this paper, we
consider $T_{e}=1$. The effect of the dust inertia is ignored for simplicity.

Energy dissipation of turbulence occurs only in the dust layer. This
assumption may be problematic for $T_{s}\gg1$, as discussed in
\S \ref{sec:coldrag}, but for simplicity, we assume that energy dissipation
occurs in $-\sqrt{2}h_{d}<z<\sqrt{2}h_{d}$. The dissipation rate of the 
energy per unit surface area and unit time is
\begin{equation}
\frac{\partial E_{\mathrm{turb}}}{\partial t}=\alpha h_{g}^{2}\Omega
_{\mathrm{K}}^{3}\int_{-\sqrt{2}h_{d}}^{\sqrt{2}h_{d}}({{\rho_{g}%
+C_{\mathrm{ene}}\rho_{d})}dz}=\alpha h_{g}^{2}\Omega_{\mathrm{K}}^{3}\Sigma
_{d,\mathrm{turb}}\ , 
\label{eq:ene_turb}%
\end{equation}
where
\begin{equation}
\Sigma_{d,\mathrm{turb}}=\Sigma_{d}\left[  {\frac{2}{{\sqrt{\pi}%
f_{\mathrm{mid}}}}+C_{\mathrm{ene}}\mathrm{erf}(1)}\right]  \ ,
\label{eq:sig_turb}%
\end{equation}
and $\mathrm{erf}(1)=0.8427$.

\section{TURBULENCE STRENGTH}
\label{sec:turbstr}

The strength of turbulence or the parameter $\alpha$ is determined by the balance
between the energy supply rate ($\partial E_{g,\mathrm{drag}}/\partial t$ in
eq. [\ref{eq:ene_drag}] and $\partial E_{\mathrm{vis}}/\partial t$ in eq.
[\ref{eq:ene_vis}]) and the energy dissipation rate ($\partial
E_{\mathrm{turb}}/\partial t$ in eq. [\ref{eq:ene_turb}]). In steady
turbulence,
\begin{equation}
\frac{\partial E_{\mathrm{turb}}}{\partial t}=C_{\mathrm{eff}}\left(
\frac{\partial E_{g,\mathrm{drag}}}{\partial t}+\frac{\partial E_{\mathrm{vis}%
}}{\partial t}\right)  \ ,
\end{equation}
where the efficiency factor $C_{\mathrm{eff}}$ represents the fraction of the
released gravitational energy that is transferred to turbulence. Comparison
with the numerical simulation by JHK06 suggests that $C_{\mathrm{eff}}=0.19$
(see \S \ref{sec:comp-sim} below), and we adopt this value in this
paper. From equations (\ref{eq:etawave}), 
(\ref{eq:ene_drag}), (\ref{eq:ene_vis}), and (\ref{eq:ene_turb}), the viscosity parameter is
\begin{equation}
\alpha=2C_{\mathrm{eff}}\tilde{\eta}T_{s}\frac{{\Sigma_{d,\mathrm{drag}%
}+\Sigma_{d,\mathrm{vis}}}}{{\Sigma_{d,\mathrm{turb}}}}. \label{eq:alpha}%
\end{equation}
Note that the right-hand-side of equation (\ref{eq:alpha}) is a function
of $\alpha$ through $f_{\mathrm{mid}}$ or $\beta$ in the ``surface
densities'' (see eqs. [\ref{eq:sig_drag}], [\ref{eq:sig_vis-approx}],
[\ref{eq:sig_vis}], [\ref{eq:sig_turb}]). Before obtaining the exact
solution for $\alpha$ numerically, in the following two subsections,
approximate solutions are derived.

\subsection{Turbulence Strength in the Small-Particle Limit ($T_{s}\ll
1$)}
\label{sec:turbstr-ts}

\begin{figure}[ptb]
\epsscale{1.1} \plotone{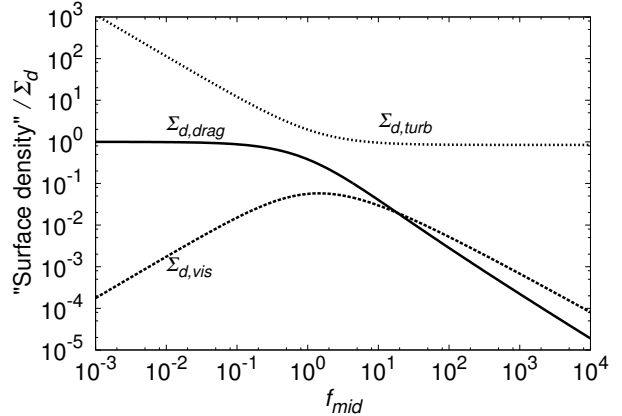}\caption{ ``Surface densities'' $\Sigma
_{d,\mathrm{drag}}$, $\Sigma_{d,\mathrm{vis}}$, and $\Sigma_{d,\mathrm{turb}}$
in the small particle limit $T_{s} \ll1$. The ``surface densities'' are
normalized by the dust surface density $\Sigma_{d}$, and are plotted as
functions of the midplane dust-to-gas ratio, $f_{\mathrm{mid}}$.}%
\label{fig:Sigma}%
\end{figure}

In equation (\ref{eq:alpha}), the \textquotedblleft surface
densities\textquotedblright, $\Sigma_{d,\mathrm{drag}}$, $\Sigma
_{d,\mathrm{vis}}$, and $\Sigma_{d,\mathrm{turb}}$ represent the effective
density of the dust that contributes to the liberation of the gravitational energy
or dissipation in the turbulence. The \textquotedblleft surface
densities\textquotedblright\ are functions of the stopping time $T_{s}$ and
the midplane dust-to-gas ratio $f_{\mathrm{mid}}$. For small dust particles
($T_{s}\ll1$), the \textquotedblleft surface densities\textquotedblright\ in
equations (\ref{eq:sig_drag}), (\ref{eq:sig_vis}), (or approximated eqs.
[\ref{eq:sig_drag2}], [\ref{eq:sig_vis-approx}]), and (\ref{eq:sig_turb})
depend on $T_{s}$ only through $f_{\mathrm{mid}}\approx Z_{\mathrm{disk}}(T_{s}/\alpha)^{1/2}$. Figure \ref{fig:Sigma} shows how the \textquotedblleft
surface densities\textquotedblright\ vary with $f_{\mathrm{mid}}$ for small
particles ($T_{s}\ll1$). When the midplane dust-to-gas ratio is much smaller
than unity, the liberation of the gravitational energy mainly comes from the
dust accreting due to individual drag, while for $f_{\mathrm{mid}}\gg1$, the
collective drag dominates the energy liberation. For the energy dissipated in
turbulence, if $f_{\mathrm{mid}}\ll1$, the \textquotedblleft surface
density\textquotedblright\ $\Sigma_{d,\mathrm{turb}}$ is the column density of
the gas in the dust layer and is inversely proportional to $f_{\mathrm{mid}}$. If the dust layer is thin enough that $f_{\mathrm{mid}}\gg1$, then
$\Sigma_{d,\mathrm{turb}}$ is simply given by the dust surface density
$\Sigma_{d}$.

Consider how the turbulent parameter $\alpha$ given by equation
(\ref{eq:alpha}) depends on the stopping time $T_{s}$ and the disk metallicity $Z_{\mathrm{disk}}$. In the following discussion, we assume small particles, such that $T_{s}\ll1$. As discussed
above, for $T_{s}\ll1$, the \textquotedblleft surface
densities\textquotedblright\ depend on $T_{s}$ only through $f_{\mathrm{mid}}\approx Z_{\mathrm{disk}}(T_{s}/\alpha)^{1/2}$, and consequently equation
(\ref{eq:alpha}) determines the value of $\alpha/T_{s}$. Thus, in the limit of
$T_{s}\ll1$, it is seen that $\alpha\propto T_{s}$. We derive an approximate
expression of $\alpha$ for the small and large limits of $f_{\mathrm{mid}}$. For
$f_{\mathrm{mid}}\ll1$, Figure \ref{fig:Sigma} (and eqs. [\ref{eq:sig_drag2}%
], [\ref{eq:sig_vis}], and [\ref{eq:sig_turb}]) show that ${\Sigma
_{d,\mathrm{drag}}}={\Sigma_{d}}$, $\Sigma_{d,\mathrm{vis}} \sim
f_{\mathrm{mid}}\Sigma_{d,\mathrm{drag}}\ll{\Sigma_{d,\mathrm{drag}}}$, 
and $\Sigma_{d,\mathrm{turb}}=2\Sigma_{d}/(\sqrt{\pi}f_{\mathrm{mid}})$. Thus,
the ratio of the ``surface densities'' in equation 
(\ref{eq:alpha}) reduces to $({\Sigma_{d,\mathrm{drag}}+\Sigma_{d,\mathrm{vis}%
})/\Sigma_{d,\mathrm{turb}}=}\sqrt{\pi}f_{\mathrm{mid}}/2$. For
$f_{\mathrm{mid}}\gg1$, though $\Sigma_{d,\mathrm{drag}}$ is smaller than
$\Sigma_{d,\mathrm{vis}}$, it still makes a contribution to the energy
liberation. We fit the functional form of $({\Sigma_{d,\mathrm{drag}}%
+\Sigma_{d,\mathrm{vis}})/\Sigma_{d,\mathrm{turb}}}$ for $10^{2}%
<f_{\mathrm{mid}}<10^{4}$ by a power law form, $({\Sigma_{d,\mathrm{drag}%
}+\Sigma_{d,\mathrm{vis}})/\Sigma_{d,\mathrm{turb}}}\approx\tilde{\Sigma
}_{d,0}f_{\mathrm{mid}}^{-\delta}$, where $\delta=0.94$, and $\tilde{\Sigma
}_{d,0}=0.71$. Using the above expressions, equation (\ref{eq:alpha}) becomes%
\begin{equation}
\alpha=\left\{
\begin{array}
[c]{ccc}%
\left(  \pi^{1/2}C_{\mathrm{eff}}\tilde{\eta}Z_{\mathrm{disk}}\right)
^{\frac{2}{3}}T_{s} & \mathrm{for} & Z_{\mathrm{disk}}\ll\sqrt{C_{\mathrm{eff}%
}\tilde{\eta}}\\
\left(  2\tilde{\Sigma}_{0}C_{\mathrm{eff}}\tilde{\eta}
Z_{\mathrm{disk}}^{-\delta} \right)  ^{\frac{2}{2-\delta}}T_{s} & \mathrm{for} &
Z_{\mathrm{disk}}\gg\sqrt{C_{\mathrm{eff}}\tilde{\eta}}%
\end{array}
\right.  , 
\label{eq:alpha1}%
\end{equation}
and the midplane dust-to-gas ratio (eq.[\ref{eq:fdmid}]) is
\begin{equation}
{f_{\mathrm{mid}}}=\left\{
\begin{array}
[c]{ccc}%
\left(  \frac{Z_{\mathrm{disk}}^{2}}{\pi^{1/2}C_{\mathrm{eff}}\tilde{\eta}}\right)
^{\frac{1}{3}} & \mathrm{for} & Z_{\mathrm{disk}}\ll\sqrt{C_{\mathrm{eff}%
}\tilde{\eta}}\\
\left(  \frac{Z_{\mathrm{disk}}^{2}}{2\tilde{\Sigma}_{0}C_{\mathrm{eff}}\tilde{\eta}}\right)
^{\frac{1}{2-\delta}} & \mathrm{for} & Z_{\mathrm{disk}}\gg\sqrt
{C_{\mathrm{eff}}\tilde{\eta}}%
\end{array}
\right.  \ . 
\label{eq:fdmid1}%
\end{equation}
The scale height of the dust layer is%
\begin{equation}
h_{d}=\left\{
\begin{array}
[c]{ccc}%
\left(  \pi^{1/2}C_{\mathrm{eff}}\tilde{\eta}Z_{\mathrm{disk}}\right)
^{\frac{1}{3}}h_{g} & \mathrm{for} & Z_{\mathrm{disk}}\ll\sqrt{C_{\mathrm{eff}%
}\tilde{\eta}}\\
\left( 2\tilde{\Sigma}_{0}C_{\mathrm{eff}}\tilde{\eta}Z_{\mathrm{disk}}^{-\delta}
\right)^{\frac{1}{2-\delta}} h_{g} & \mathrm{for} &
Z_{\mathrm{disk}}\gg\sqrt{C_{\mathrm{eff}}\tilde{\eta}}%
\end{array}
\right.  ~. 
\label{eq:hd3}%
\end{equation}
In the above approximate expressions (\ref{eq:alpha1})-(\ref{eq:hd3}), the
upper line represents $f_{\mathrm{mid}}\ll1$ and the lower line represents
$f_{\mathrm{mid}}\gg1$. The transition of these expressions occurs at
$f_{\mathrm{mid}}\sim1$, which corresponds to $Z_{\mathrm{disk}}\sim
\sqrt{C_{\mathrm{eff}}\tilde{\eta}}$, and a maximum value of $\alpha$,
\begin{equation}
\alpha_{\max}\sim C_{\mathrm{eff}}{\tilde{\eta}}T_{s}\ . 
\label{eq:alpha2}%
\end{equation}

\begin{figure}[ptb]
\epsscale{1.1} \plotone{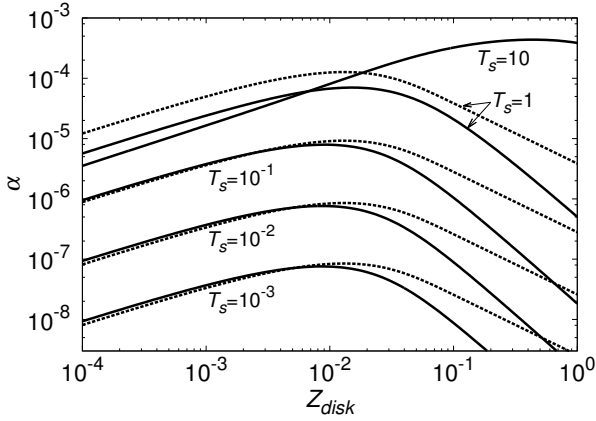}\caption{The turbulent viscosity parameter
$\alpha$ for various values of the stopping time $T_{s}$ plotted as a
function of the disk metallicity $Z_{\mathrm{disk}}$. The solid lines
show $\alpha$ derived from our model. The dashed lines show $\alpha$ from the S98
model. We adopt an efficiency parameter of the energy supply,
$C_{\mathrm{eff}}=0.19$, for our model, and the critical Richardson
number $\mathrm{Ri}=0.8$ for the S98 model.}
\label{fig:alpha}%
\end{figure}

\begin{figure}[ptb]
\epsscale{1.1} \plotone{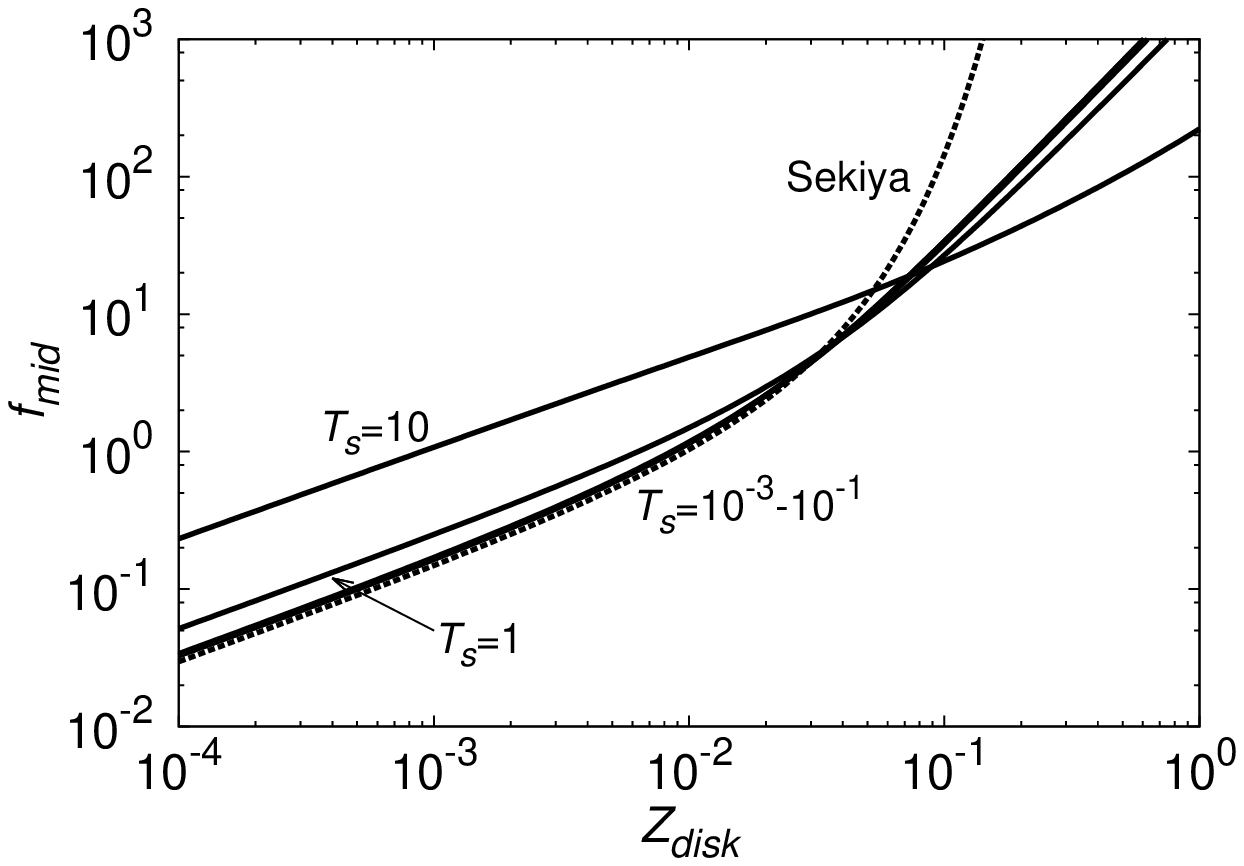}\caption{The midplane dust-to-gas ratio
$f_{\mathrm{mid}}$. The solid lines are calculated for various values of the
stopping time $T_{s}$ using our model. The dashed line is taken from the S98 model.}%
\label{fig:fdmid}%
\end{figure}

\subsection{Turbulence Strength for a Small Dust-to-Gas
Ratio($f_{\mathrm{mid}}\ll1$)}
\label{sec:turbstr-fm}

We derive the approximate expression of the turbulent
strength $\alpha$ as a function of $T_{s}$ and $Z_{\mathrm{disk}}$ in the
limit of small dust-to-gas ratio at the midplane, $f_{\mathrm{mid}}\ll1$,
but without assuming $T_{s}\ll1$. For $f_{\mathrm{mid}}\ll1$, the ``surface
densities'' (eqs. [\ref{eq:sig_drag}], [\ref{eq:sig_vis}], and
[\ref{eq:sig_turb}]) reduce to $\Sigma_{d,\mathrm{drag}}=\frac{1}{2}\Sigma
_{d}(T_{s}^{2}+2)/(T_{s}^{2}+1)^{2}$, $\Sigma_{d,\mathrm{vis}}\sim
f_{\mathrm{mid}}\Sigma_{d,\mathrm{drag}}\ll\Sigma_{d,\mathrm{drag}}$, and
$\Sigma_{d,\mathrm{turb}}=2\Sigma_{d}/(\sqrt{\pi}f_{\mathrm{mid}})$. Thus,
the approximate expression of equation (\ref{eq:alpha}) becomes%
\begin{equation}
\alpha=\left[  \frac{\sqrt{\pi}}{2}C_{\mathrm{eff}}\tilde{\eta}%
Z_{\mathrm{disk}}\sqrt{\frac{1+2T_{s}}{1+T_{s}}}\frac{T_{s}^{2}+2}{\left(
T_{s}^{2}+1\right)  ^{2}}\right]  ^{\frac{2}{3}}T_{s}~. 
\label{eq:alpha3}%
\end{equation}
The midplane dust-to-gas ratio and the scale height of the dust layer are,
respectively,%
\begin{equation}
f_{\mathrm{mid}}=\left[  \frac{2Z_{\mathrm{disk}}^{2}}{\sqrt{\pi
}C_{\mathrm{eff}}\tilde{\eta}}\frac{1+2T_{s}}{1+T_{s}}\frac{\left(  T_{s}%
^{2}+1\right)  ^{2}}{T_{s}^{2}+2}\right]  ^{\frac{1}{3}}~,
\end{equation}
and%
\begin{equation}
h_{d}=\left[  \frac{\sqrt{\pi}}{2}C_{\mathrm{eff}}\tilde{\eta}%
Z_{\mathrm{disk}}\frac{1+T_{s}}{1+2T_{s}}\frac{T_{s}^{2}+2}{\left(  T_{s}%
^{2}+1\right)  ^{2}}\right]  ^{\frac{1}{3}}h_{g}~.
\end{equation}

\subsection{Turbulence Strength and Dust Layer Thickness}

The properties of the turbulent viscosity parameter $\alpha$ for $T_{s}\ll1$ or
$f_{\mathrm{mid}}\ll1$ are described in the last two subsections. For general
$T_{s}$ and $f_{\mathrm{mid}}$, equation (\ref{eq:alpha}) must be
solved numerically. We use the van Wijngaarden-Dekker-Brent method (Press et
al. 1986) to obtain $\alpha$ from equation (\ref{eq:alpha}). In Figure
\ref{fig:alpha}, the solid lines represent the variation of $\alpha$ with the disk metallicity $Z_{\mathrm{disk}}$ for various values of the particle
stopping time $T_{s}$. It is seen that, for small dust particles ($T_{s}%
\la1$), $\alpha$ is proportional to $T_{s}$, as discussed in
\S \ref{sec:turbstr-ts}. The maximum $\alpha_{\mathrm{max}}\sim
C_{\mathrm{eff}}\tilde{\eta}T_{s}$ appears at $Z_{\mathrm{disk}}\sim
\sqrt{C_{\mathrm{eff}}\tilde{\eta}}\sim10^{-2}$ (and $f_{\mathrm{mid}}\sim
1$) for $T_{s}\la1$. The turbulence weakens as $Z_{\mathrm{disk}}$
deviates from $\sqrt{C_{\mathrm{eff}}\tilde{\eta}}$ ($\alpha\propto
Z_{\mathrm{disk}}^{2/3}$ for $Z_{\mathrm{disk}}\la\sqrt{C_{\mathrm{eff}%
}\tilde{\eta}}$, and $\alpha\propto Z_{\mathrm{disk}}^{-1.8}$ for
$Z_{\mathrm{disk}}\ga\sqrt{C_{\mathrm{eff}}\tilde{\eta}}$, see eq.
[\ref{eq:alpha1}]). For particles of $T_{s}>1$, $\alpha$ peaks
at a larger $Z_{\mathrm{disk}}$ (or $f_{\mathrm{mid}}$).

\begin{figure}[ptb]
\epsscale{1.1} \plotone{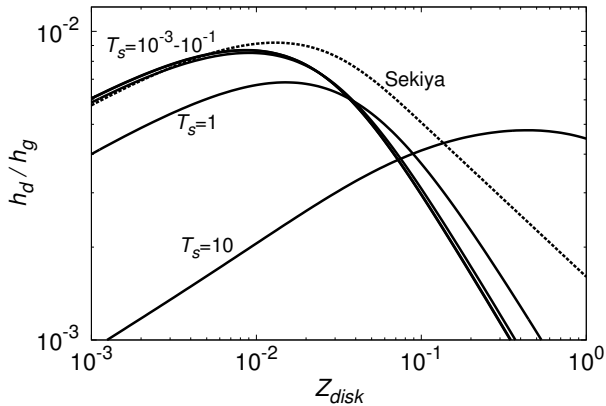}\caption{The scale height of the dust layer
$h_{d}$ normalized by the gas scale height $h_g$. The solid lines are
  calculated for various values of the stopping time $T_{s}$ by our
  model. The dashed line is taken from the S98 model, which is
  calculated using equation (\ref{eq:hd-sek}).}
\label{fig:hd}%
\end{figure}

Figure \ref{fig:fdmid} shows the variation in the midplane dust-to-gas ratio
$f_{\mathrm{mid}}$ with the disk metallicity $Z_{\mathrm{disk}}$. For small particles ($T_{s}\la1$), $\alpha$ is
proportional to $T_{s}$, and the midplane dust-to-gas ratio
$f_{\mathrm{mid}}\propto(T_{s}/\alpha)^{1/2}$ does not depend on
$T_{s}$. For a small dust-to-gas ratio ($f_{\mathrm{mid}}\ll1$),
$f_{\mathrm{mid}}\propto Z_{\mathrm{disk}}^{2/3}$, and for $f_{\mathrm{mid}}\gg1$,
$f_{\mathrm{mid}}\propto Z_{\mathrm{disk}}^{1.9}$. Note that if the disk metallicity is larger than $0.2$, then the midplane dust-to-gas ratio
becomes as large as $100$, which is high enough to induce a gravitational
instability in the dust layer. This result is consistent with the previous
works by, e.g., S98. Figure \ref{fig:hd} shows that the dust layer thickness $h_{d}$
is of the order of $10^{-3}-10^{-2}$ times the thickness of the gas disk.

\section{COMPARISONS WITH PREVIOUS STUDIES}
\label{sec:comppre}

In this section, the dust layer model that is based on the energetics of
dust accretion is compared with results from previous work. Previous studies 
have analyzed the detailed physics of the dust layer, including the
onset of KH or streaming instabilities, and several stabilizing effects, such as
Keplerian shear, using both analytical and numerical approaches. It is of
interest to determine which properties of the dust layer are reproduced by our
model and which are missing.

\subsection{Comparison with Previous Analytical Studies}

\subsubsection{Comparison with Sekiya (1998)}

S98 analytically solved the density structure of the dust layer, assuming that
the layer structure was adjusted to be marginally unstable to KH instability.
To obtain the density structure, S98 assumed that the Richardson number remained 
close to the critical value for instability throughout the dust
layer. S98 considered small dust particles that were tightly coupled to the gas 
($T_{s}\ll1$), and treated the gas and the dust as a single fluid. Then, the
density structure of the dust layer was calculated for various values of the
disk metallicity $Z_{\mathrm{disk}}$. In S98, the dust layer
structure was determined by the argument for the stability of a
stratified fluid, not 
by the balance between sedimentation and diffusion of the dust. However, it is
possible to interpret the result of S98 as follows: The turbulent
strength in the dust layer is adjusted such that the turbulent diffusion of
the dust maintains the marginally unstable density structure. Using this
interpretation, we calculate the effective value of the turbulent diffusion
parameter $\alpha$ as a function of $Z_{\mathrm{disk}}$ for $T_{s}\ll1$.

From equation (22) of S98, the midplane dust-to-gas ratio,
$f_{\mathrm{mid}}$, is related to the disk metallicity,
$Z_{\mathrm{disk}}$, as
\begin{eqnarray}
{Z_{\mathrm{disk}}} &=& \sqrt{\frac{{2\mathrm{{Ri}}}\tilde{\eta}}{\pi}} \left\{
\ln\left[  {\left(  {1+\sqrt{1-{{\left(  {\frac{1}{{1+{f_{\mathrm{mid}}}}}
}\right)  }^{2}}}}\right)  \left(  {1+{f_{\mathrm{mid}}}}\right)  }\right]
\right. \nonumber \\
& &\left. -\sqrt{1-{{\left(  {\frac{1}{{1+{f_{\mathrm{mid}}}}}}\right)  }^{2}}}
\right\} \ , 
\label{eq:fd-sek}%
\end{eqnarray}
where the Richardson number $\mathrm{{Ri}}$ is constant
throughout the dust layer, and the self-gravity of the dust layer is neglected.
The dust density distribution is given by
\begin{equation}
\frac{{{\rho_{d}}}}{{{\rho_{g}}}}={\left[  {\frac{{{z^{2}}}}{{\mathrm{{Ri}%
}{\tilde{\eta}h}}_{g}^{2}}+{{\left(  {\frac{1}{{1+{f_{\mathrm{mid}}}}}%
}\right)  }^{2}}}\right]  ^{-\frac{1}{2}}}-1\ . 
\label{eq:dendist-sek}%
\end{equation}
The half-thickness of the dust layer, $z_{d}$, at which the dust density
becomes zero, is given by
\begin{equation}
z_{d}=\sqrt{\mathrm{{Ri}}\tilde{\eta}\left[  {1-(1+f_{\mathrm{mid}})}%
^{-2}\right]  }\mathrm{\ }h_{g}~. 
\label{eq:zd}%
\end{equation}
To compare with the scale height of Gaussian distribution, $h_{d}$, we define
the scale height of the dust distribution as the vertical dispersion of the
dust particles,
\begin{equation}
h_{d,\mathrm{{Sek}}}^{2}=\frac{\int_{0}^{z_{d}}z^{2}\rho_{d}dz}{\int
_{0}^{z_{d}}\rho_{d}dz}\ . 
\label{eq:hd-sek}%
\end{equation}
The turbulent parameter $\alpha_{\mathrm{{Sek}}}$ is estimated from equation
(\ref{eq:hd}), using $T_{s}\ll1$, as
\begin{equation}
\alpha_{\mathrm{{Sek}}}=\left(  \frac{h_{d,\mathrm{{Sek}}}}{h_{g}}\right)
^{2}T_{s}\ . 
\label{eq:alpha-sek}%
\end{equation}
For $f_{\mathrm{mid}}\ga1$, equation (\ref{eq:zd}) reduces to $z_{d}%
\approx(\mathrm{{Ri}}\tilde{\eta})^{1/2}h_{g}$. Substituting this expression
into $h_{d,\mathrm{Sek}}\approx z_{d}$ of equation (\ref{eq:alpha-sek}) gives
\begin{equation}
\alpha_{\mathrm{{Sek}}}\approx\mathrm{{Ri}}{\tilde{\eta}T_{s}}%
\ \ \ \mathrm{for}\ \ \ f_{\mathrm{mid}}\ga1\ , 
\label{eq:alpha-sek2}%
\end{equation}
which can be compared with equation (\ref{eq:alpha2}). For $f_{\mathrm{mid}}\ll1$, equation (\ref{eq:fd-sek}) is approximated in the lowest order of
$f_{\mathrm{mid}}$ as
\begin{equation}
{Z_{\mathrm{disk}}}=\frac{4}{3}\sqrt{\frac{{\mathrm{{Ri}}}\tilde{\eta}}{\pi}%
}\left[  {f_{\mathrm{mid}}^{3/2}+O(f_{\mathrm{mid}}^{5/2})}\right]  .
\end{equation}
Thus, the midplane dust-to-gas ratio is
\begin{equation}
f_{\mathrm{mid}}=\left(  { \frac{9}{16}\frac
{\pi}{{\mathrm{{Ri}}\tilde{\eta}}}Z_{\mathrm{disk}}^{2}}\right)^{\frac
{1}{3}}\ . 
\label{eq:fdmid-sek}%
\end{equation}
For $f_{\mathrm{mid}}\ll1$ and $z \ll h_g$, the dust density distribution
(eq.[\ref{eq:dendist-sek}]) is
\begin{equation}
\frac{\rho_{d}}{\rho_{g}}
 \approx{f_{\mathrm{mid}}}\left[  {1-\frac{{1}}{{2\mathrm{{Ri}}%
{\tilde{\eta}f_{\mathrm{mid}}}}}\left(  \frac{z}{h_{g}}\right)  }^{2}%
 \right]  \ .
\end{equation}
To the order of $(z/h_{d})^{2}$, this distribution is approximated by the
Gaussian distribution $f_{\mathrm{mid}}
\exp[-z^{2}/(2h_{d,\mathrm{Sek}}^{2})] \approx f_{\mathrm{mid}}
    [1-z^{2}/(2h_{d,\mathrm{Sek}}^{2})]$, and its scale height $h_{d,\mathrm{Sek}}$ is 
\begin{equation}
h_{d,\mathrm{Sek}}\approx\sqrt{\mathrm{{Ri}}\tilde{\eta}{f_{\mathrm{mid}}}%
}h_{g}\approx{\left[  {\frac{{\mathrm{{3Ri}}}}{4}{\pi^{1/2}}{Z_{\mathrm{disk}}}\tilde{\eta}}\right]  ^{\frac{1}{3}}}{h_{g}}\ .
\end{equation}
The turbulent parameter $\alpha_{\mathrm{Sek}}$ of equation
(\ref{eq:alpha-sek}) is
\begin{equation}
\alpha_{\mathrm{Sek}}\approx{\left[  {\frac{{\mathrm{{3Ri}}}}{4}{\pi^{1/2}%
}{Z_{\mathrm{disk}}}\tilde{\eta}}\right]  ^{\frac{2}{3}}}{T_{s}%
}\ \ \ \ \mathrm{for}\ \ \ f_{\mathrm{mid}}\ll1\ . 
\label{eq:alpha-sek3}%
\end{equation}
Comparing this expression to equation (\ref{eq:alpha1}) provides a relationship
between the energy supply efficiency to turbulence, $C_{\mathrm{eff}}$, and
the critical Richardson number, $\mathrm{Ri}$, as ${C_{\mathrm{eff}}%
}=(3/4)\mathrm{{Ri}}$. In the above discussion on $\alpha_{\mathrm{Sek}}$, we
used the dust layer thickness $h_{d,\mathrm{Sek}}$ derived from the Gaussian
fit. However, the thickness defined by the vertical dispersion
(eq.[\ref{eq:hd-sek}]) in the small $f_{\mathrm{mid}}$ limit is $0.63$ times
thinner than that defined by the Gaussian fit. Thus, the energy supply
efficiency is $0.4=0.63^{2}$ times smaller than the above estimate and thus
$C_{\mathrm{eff}}\approx0.3\mathrm{{Ri}}$. The relationship between
$C_{\mathrm{eff}}$ and $\mathrm{Ri}$ can also be derived by comparing the
expressions for $f_{\mathrm{mid}}$ (eqs. [\ref{eq:fdmid1}] and
[\ref{eq:fdmid-sek}]) as ${C_{\mathrm{eff}}}=16/(9\pi^{3/2})\mathrm{{Ri}%
}\approx0.3\mathrm{{Ri}}$. We compare our result with the numerical simulation
developed by JHK06, which will be discussed in \S \ref{sec:comp-sim}. The comparison
with this simulation suggests a slightly smaller energy supply efficiency,
\begin{equation}
C_{\mathrm{eff}}\approx0.24\mathrm{{Ri}}\ . 
\label{eq:ceff-ri}%
\end{equation}

The turbulent parameter $\alpha_{\mathrm{Sek}}$ given by equation
(\ref{eq:alpha-sek}) is numerically calculated and is plotted using dashed lines
in Figure \ref{fig:alpha}. The critical Richardson number $\mathrm{Ri}=0.8$ is
adopted to fit the simulation results provided by JHK06. For $Z_{\mathrm{disk}}\ll1$
and $T_{s}\la1$, the solid and dashed lines agree with each other very
well. This is expected from the above discussion that the dependence of
$\alpha$ and $\alpha_{\mathrm{Sek}}$ on $Z_{\mathrm{disk}}$ and $T_{s}$ for
small $f_{\mathrm{mid}}$ are the same (eqs. [\ref{eq:alpha1}] and
[\ref{eq:alpha-sek3}]). For $Z_{\mathrm{disk}}\ga 10^{-2}$, a deviation
between the solid and dashed lines appears, and the difference increases
with $Z_{\mathrm{disk}}$. We note that for $Z_{\mathrm{disk}} \ga 0.1$,
the S98 density distribution shows a cusp at the midplane, which is
a significant difference from the Gaussian distribution of our model. In
Figure \ref{fig:fdmid}, the midplane dust-to-gas ratio of the S98 model
is plotted with a dashed line. For small $Z_{\mathrm{disk}}\la0.05$, the
result from S98 is consistent with our result (the dashed line coincides
with the solid lines for $T_{s}\la1$). For 
$Z_{\mathrm{disk}}\ga0.05$, the midplane dust-to-gas ratio from the S98 model
is larger than our results. Both results are still qualitatively consistent,
indicating that, for $Z_{\mathrm{disk}}\ga0.1$, the midplane dust-to-gas
ratio becomes large enough for gravitational instability ($f_{\mathrm{mid}}\sim100$).

\subsubsection{Comparison with Michikoshi \& Inutsuka (2006)}

\begin{figure}[ptb]
\epsscale{1.1} \plotone{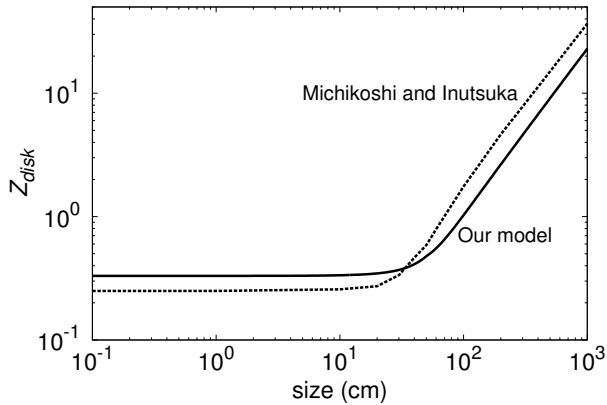}\caption{Comparison with the results in
Michikoshi \& Inutsuka (2006). The solid line shows the relationship between the
midplane dust-to-gas ratio $f_{\mathrm{mid}}$ and the particle radius $a$.
The dashed line shows the boundary at which the growth rate of the KH instability
is $\omega=0.1$ (above the dashed line, $\omega>0.1$). We use the same disk
model that is adopted in Michikoshi \& Inutsuka (2006). The dashed line is read from
Fig. 16 in Michikoshi \& Inutsuka (2006).}%
\label{fig:michi}%
\end{figure}

Michikoshi \& Inutsuka (2006) analyzed the growth rate of the KH instability
of a dust layer, taking into account the relative motion and the friction
between the dust and the gas. Their formulation does not assume tight dust-gas
coupling, and thus it can be applied to scenarios involving large dust particles ($T_{s}\gg1$),
while the vertical gravity, Coriolis force, and Keplerian shear are neglected.
The initial velocity gradient in the vertical direction is caused by the dust
inertia, using the formula provided in NSH86. The growth rate of instability has been derived
for a wide range of dust sizes $a$ and midplane dust-to-gas
ratios $f_{\mathrm{mid}}$, and is summarized in their Figure 16. They argued
that if the effect of Keplerian shear is taken into account, the line
corresponding to the growth rate $\omega=1$ in the $a$-$f_{\mathrm{mid}}$
plane would be the boundary between the stable and unstable
configurations. This marginally unstable structure of the dust layer
presented by Michikoshi \& Inutsuka (2006) can be considered an
extension of the S98 model for general sizes of the dust particles, and
provides a reference to be compared with our model.

In Figure \ref{fig:michi}, we compare our model in the $a$-$f_{\mathrm{mid}}$
plane with the line at which the growth rate $\omega$ has a constant critical
value. We adopt the critical growth rate as $\omega=0.1$, which is smaller
than the value Michikoshi \& Inutsuka (2006) suggested. The qualitative
behavior does not differ between $\omega=0.1$ and $\omega=1$. Our model well
reproduces the result of Michikoshi \& Inutsuka (2006). The agreement of our model
with the results of S98 and of Michikoshi \& Inutsuka (2006) suggests that the
onset of KH instability is controlled by the energy supply due to dust
accretion over a wide range of particle sizes, $a$, or stopping times,
$T_{s}$. It must be noted, however, that the initial state assumed in Michikoshi \&
Inutsuka (2006) is such that the vertical velocity shear appears only
in the dust layer, neglecting the velocity shear in the Ekman-like boundary
layer that may appear between the dust layer and the upper gas layer. This is
the same assumption that we adopt. As discussed in \S \ref{sec:coldrag}, this
assumption is appropriate for small dust particles ($T_{s}\la1$), while for
large particles ($T_{s}\gg1$) the vertical velocity profile of the thick turbulent 
boundary layer may quickly deviate from that given by NSH86. Though the
agreement with the result by Michikoshi \& Inutsuka (2006) shows the applicability
of our model to KH instability in the dust layer for general values of $T_{s}%
$, we should keep in mind the limitation of the models mentioned above.

\begin{figure}[ptb]
\epsscale{1.1} \plotone{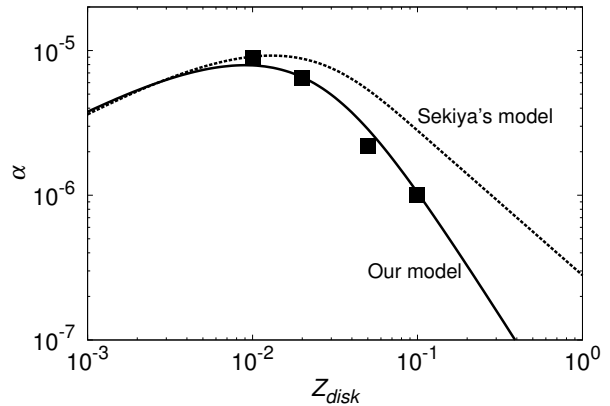}\caption{The turbulent viscosity parameter $\alpha$
estimated from our model, compared with the simulation described in JHK06. The solid
line is calculated from our model and the dashed line is taken from the S98 model. The
squares are the simulation results, $\delta_{t}$ in Table 2 of JHK06. The
stopping time $T_{s}=0.1$ is adopted both for our model and the simulation.}%
\label{fig:alpha-joh}%
\end{figure}

\subsection{Comparison with Previous Numerical Simulations}
\label{sec:comp-sim}

\subsubsection{Comparison with Johansen et al. (2006)}

Numerical simulations of KH instability induced by formation of
a dust layer were presented in JHK06. Their two-dimensional 
simulation on the $\theta z$-plane solved for
both the gas and dust motions. They obtained a quasi-steady or oscillating
density distribution of the dust in which dust settling and turbulent
diffusion balanced each other. The diffusion coefficient in the vertical
direction, $\delta_{t}$, was measured from the scale height of the dust layer
and was summarized in their Table 2. In Figure \ref{fig:alpha-joh}, the
turbulent viscosity parameter $\alpha$ of equation (\ref{eq:alpha}) (the solid
line) is compared with the diffusion coefficient measured from the simulation
(the squares). To fit the simulation result, the efficiency parameter of the
energy supply to turbulence, $C_{\mathrm{eff}}$, is set to $0.19$. We also
fit the turbulent diffusion parameter for the density distribution of S98,
$\alpha_{\mathrm{Sek}}$, to the simulation. We adopt the critical Richardson
number as $\mathrm{Ri}=0.8$ for fitting. This is slightly smaller than the
value JHK06 suggested ($\mathrm{Ri}=1.0$). It is seen that the numerical
simulation is well explained both by our model and the S98 model. This means that the
dust layer is maintained so as to provide a constant Richardson number, and in
the parameter range that JHK06 surveyed ($0.01<Z_{\mathrm{disk}}<0.1$), this
critical Richardson number does not vary significantly with the disk
metallicity $Z_{\mathrm{disk}}$. (The recent result by Lee et al. (2010), in which
they argue that $\mathrm{Ri}$ is proportional to $Z_{\mathrm{disk}}$ if the
Keplerian shear is taken into account, will be discussed later in this
subsection.) Because the efficiency parameter $C_{\mathrm{eff}}$ in our model
is proportional to the critical Richardson number, as shown in equation
(\ref{eq:ceff-ri}), the efficiency parameter $C_{\mathrm{eff}}$ is also
expected to be constant for $0.01<Z_{\mathrm{disk}}<0.1$.

\begin{figure}[ptb]
\epsscale{1.1} \plotone{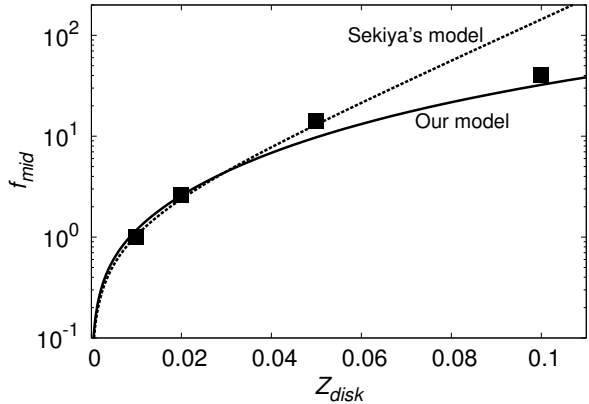}\caption{Midplane dust-to-gas ratio
$f_{\mathrm{mid}}$ derived from analytical models and from the simulation described in
JHK06. The solid line is calculated from our model and the dashed line is taken from
the S98 model. The squares are the simulation results, which are read from
Figure 13 of JHK06. The stopping time is $T_{s}=0.1$.}%
\label{fig:fdmid-joh}%
\end{figure}

\begin{figure}[ptb]
\epsscale{1.1} \plotone{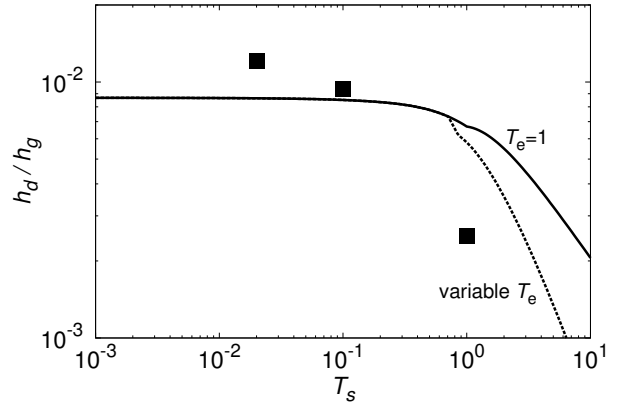}\caption{The scale heights of the dust layer $h_{d}$
estimated from our model and from the JHK06 simulation are plotted versus the
stopping time $T_{s}$. The solid line is calculated assuming that the turnover
time of the largest eddies is the Keplerian time ($T_{e}=1$). The dashed line
includes the variable $T_{e}$ calculated by equation (\ref{eq:Teddy}). The squares
are the simulation results, $\langle z^{2}\rangle^{1/2} /H$ in
Table 2 of JHK06. The disk metallicity is $Z_{\mathrm{disk}}=10^{-2}$.}%
\label{fig:hd-ts-joh}%
\end{figure}

A comparison of the midplane dust-to-gas ratio between our model and the
simulation is shown in Figure \ref{fig:fdmid-joh}. The simulation is well
fitted by our model (the solid line) and also by the S98 model (the dashed
line). For $Z_{\mathrm{disk}}=0.1$, the S98 model predicts larger
$f_{\mathrm{mid}}$ than the simulation. This may be because the simulation
does not have enough resolution to resolve the density structure around the
midplane at which the S98 model expects a rapid increase in the dust density. If
this is the case, it is difficult to judge whether our model or the model presented by S98
is the best fit with the simulation.

Figure \ref{fig:hd-ts-joh} shows the variation in dust layer thickness with
$T_{s}$ from our model and compares this variation with the simulation
results. In plotting Figure \ref{fig:alpha-joh} in the last paragraph,
the turbulent viscosity parameter $\alpha$ was calculated 
from equation (\ref{eq:hd}) using the dust layer thickness $h_{d}$ measured
from the simulation. However, it is not clear if equation (\ref{eq:hd}) is
still valid for large $T_{s}$ because this equation assumes that the turnover time
of the largest eddies is equal to the Keplerian time (YL07), and as discussed below
this assumption may not be appropriate for large $T_{s}$. Thus, in Figure
\ref{fig:hd-ts-joh}, the dust layer thickness is plotted directly without
transferring to $\alpha$. In our model, the dust layer thickness $h_{d}$ is
constant for $T_{s}\la1$ and decreases slowly with $T_{s}$ for $T_{s}\ga1$
(the solid line), although in the simulation it decreases more rapidly with
$T_{s}$. The discrepancy between our model and the simulation is apparent for
$T_{s}=1$. The difference is as large as a factor of 3, but it causes an order of magnitude
discrepancy in $\alpha\propto h_{d}^{2}$. This discrepancy suggests that our
model predicts turbulent diffusion that is too large for $T_{s}\ga1$. In fact, for
large $T_{s}$ in our model, the dust layer thickness becomes smaller than the
size of the largest eddies, contradicting our assumption that the turbulent
layer coincides with the dust layer (see, however, the discussion in
\S \ref{sec:coldrag} on the validity of this assumption, and see also eq. [56]
of YL07 for a possible physical reason for $l_{g,\mathrm{eddy}} > h_{d}$).
From equation (\ref{eq:hd}) and $l_{g,\mathrm{eddy}}=\alpha^{1/2}h_{g}$, the
condition for the dust layer thickness to be larger than the largest eddy size
($h_{d}>l_{g,\mathrm{eddy}}$) is $T_{s}<1/\sqrt{2}$. Thus, for $T_{s}%
>1/\sqrt{2}$, our model results in a dust layer that is too thick (or an eddy
size that is too small). One possible remedy for this situation is to remove the assumption that
the turnover time of the largest eddies is equal to the Keplerian time. By
introducing a parameter $\xi$ ($0\leq \xi \leq1/2$), the largest eddy size and the
velocity are expressed as $l_{g,\mathrm{eddy}}=\alpha^{(1/2)+\xi}h_{g}$ and
$u_{g,\mathrm{eddy}}=\alpha^{(1/2)-\xi}c_{s}$. The non-dimensional turnover time
of the largest eddies is $T_{e}=(l_{g,\mathrm{eddy}}/u_{g,\mathrm{eddy}%
})\Omega_{\mathrm{K}}=\alpha^{2\xi}$. The dust layer thickness for $T_{e}\neq1$
is provided by equation (21) of YL07,
\begin{equation}
{h_{d}}=\sqrt{\frac{\alpha}{{{T_{s}}}}}{\left(  {1+\frac{{{T_{s}}{T_{e}}^{2}}%
}{{{T_{s}}+{T_{e}}}}}\right)  ^{-1/2}}{h_{g}}\ .
\end{equation}
For $T_{s}>1/2$, we impose the condition that the largest eddy size
$l_{g,\mathrm{eddy}}=\alpha^{(1/2)+\xi}h_{g}$ is equal to the above $h_{d}$.
This condition determines the turnover time by the equation
\begin{equation}
{T_{s}}{T_{e}}({T_{s}}+{T_{e}}+{T_{s}}{T_{e}}^{2})-{T_{s}}-{T_{e}}=0\ .
\label{eq:Teddy}%
\end{equation}
Using this $T_{e}$ (or $\xi$), the turbulent viscosity parameter $\alpha$ and
the dust layer thickness $h_{d}$ are recalculated and plotted with the dashed
line in Figure \ref{fig:hd-ts-joh}. As expected, the introduction of a new
parameter ${T_{e}}$ suppresses the largest eddy size and the turbulent
diffusion of dust particles for $T_{s}>1/\sqrt{2}$, improving the comparison with the
simulation result. However, we need to determine whether the turnover
time of the largest eddies, $T_{e}\Omega_{\mathrm{K}}^{-1}=\alpha^{2\xi}%
\Omega_{\mathrm{K}}^{-1}$, in the simulation presented in JHK06 for $T_{s}=1$ is
actually smaller than the Keplerian time, as equation (\ref{eq:Teddy})
predicts. Cuzzi et al. (1993) argue that, based on laboratory
measurements, the eddy turnover time in an Ekman layer would be
smaller than the Keplerian time by a factor $20-80$. The turbulent
layer possibly behaves as an Ekman layer for $T_s >1$, as discussed in
\S\ref{sec:coldrag}. We are currently performing numerical
simulations using the same conditions as JHK06 to investigate the
turbulence for $T_{s} \ga 1$ in more detail (Ishitsu et al., in preparation).

Lee et al. (2010) performed a three-dimensional numerical simulation of the
onset of KH instability. They solved simplified equations in which the
dust and the gas were treated as a single fluid, but they included the effect of
the Keplerian shear in the radial direction. They found that the radial shear
stabilizes the KH instability, and the critical Richardson number for instability
is not always the standard value, $0.25$, but can be much smaller if the
stabilizing effect of the radial shear is significant (see also Ishitsu \&
Sekiya 2003). Equation (32) in Lee et al. (2010) shows that the ratio of the
stabilizing effect by the radial shear to the destabilizing effect by the
vertical shear is proportional to $\mathrm{Ri}(1+f_{\mathrm{mid}})/f_{\mathrm{mid}}\approx\mathrm{Ri}/f_{\mathrm{mid}}$ for
$f_{\mathrm{mid}}\ll1$, and thus the critical Richardson number should scale
as $\mathrm{Ri}\propto f_{\mathrm{mid}}$. In our model, we assume that the
energy supply efficiency $C_{\mathrm{eff}}$, which is proportional to the
critical Richardson number (eq.[\ref{eq:ceff-ri}]), is constant. However, the
simulation in Lee et al. (2010) suggests that
$C_{\mathrm{eff}}$ should also be proportional to $f_{\mathrm{mid}}$
(even for $f_{\mathrm{mid}} \ga 1$). If this is the case,
dependence of $f_{\rm mid}$ on $Z_{\rm disk}$ would be milder than
that shown in Figure \ref{fig:fdmid} ($f_{\rm mid} \propto Z_{\rm
disk}^{1/2}$ for $f_{\rm mid} \ll 1$ and $f_{\rm mid} \propto Z_{\rm
disk}^{0.97}$ for $f_{\rm mid} \gg 1$).
The effect of the radial shear was not included
in the simulation presented in JHK06, with which we compared our model
in detail, because their simulation was two-dimensional in the $\theta
z$-plane. Extending JHK06 to three dimensions and including the radial
shear effect are crucial to determining how the energy supply efficiency
$C_{\mathrm{eff}}$ behaves as $f_{\mathrm{mid}}$ varies.

\subsubsection{Comparison with Bai \& Stone (2010)}
\label{sec:BS10}

\begin{figure}[ptb]
\epsscale{1.1} \plotone{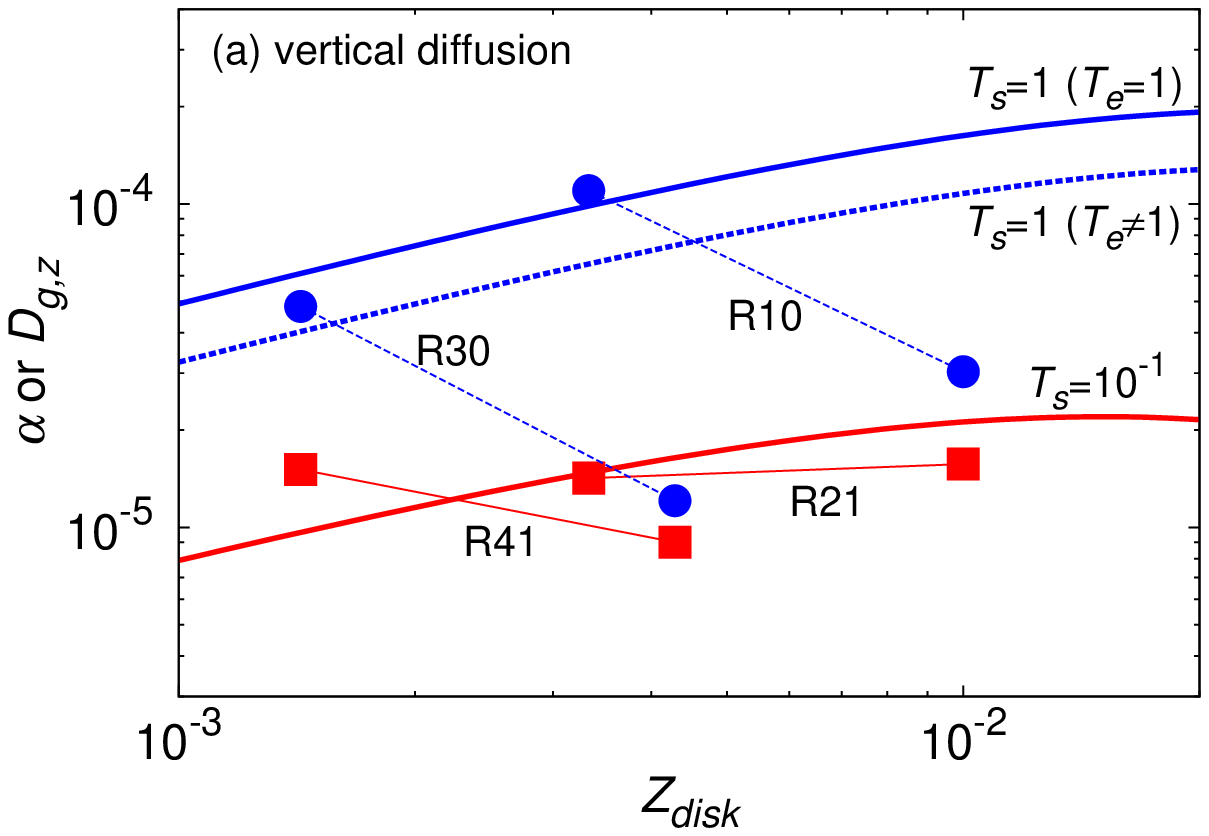} \plotone{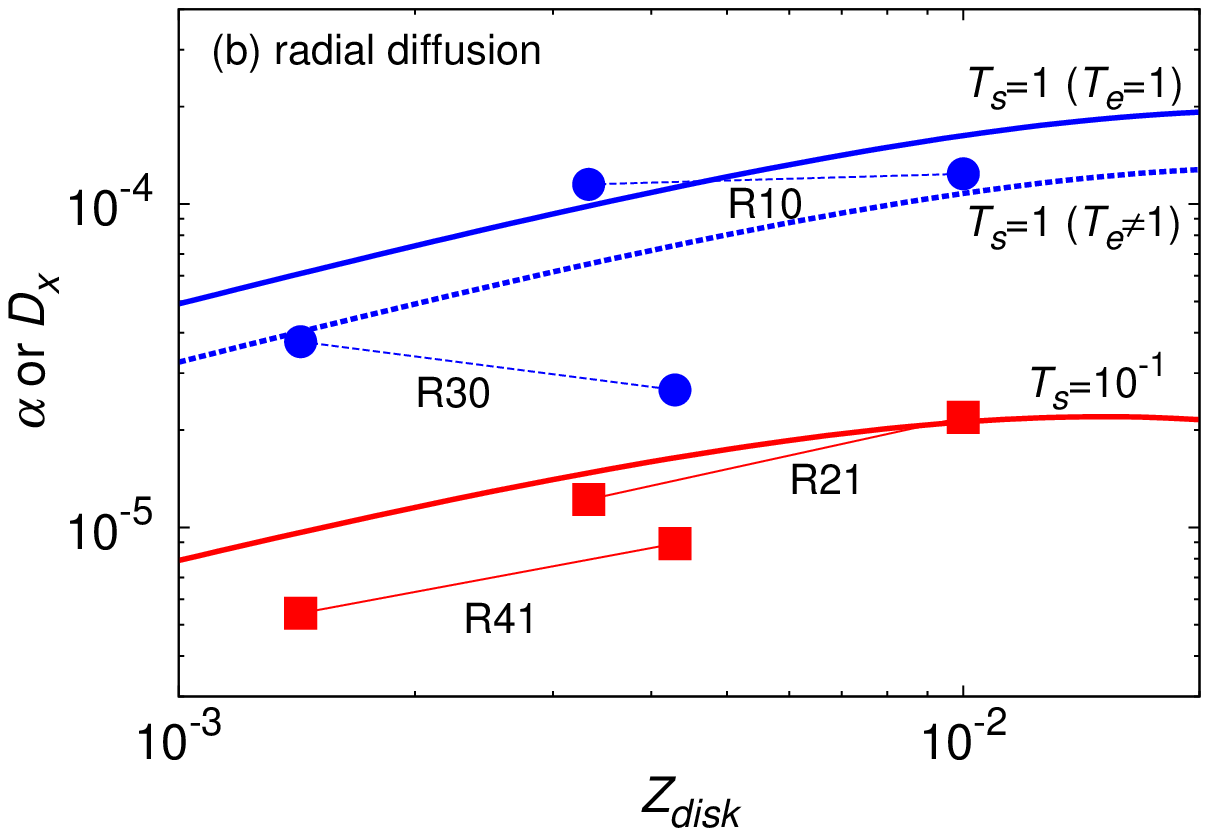} 
\caption{The turbulent
viscosity parameter $\alpha$ estimated from our model for $T_{s}=10^{-1}$ and 1
is compared with the simulation presented in Bai \& Stone (2010). The solid line is
calculated assuming the turnover time of the largest eddies is the Keplerian
time ($T_{e}=1$). The dashed line includes the variable $T_{e}$ calculated from
equation (\ref{eq:Teddy}). $(a)$ Comparison with the vertical diffusion
coefficient, $D_{g,z}(3\mathrm{D})$, measured in the simulation (Table 2 of
Bai \& Stone 2010). The squares and circles are the simulation results, 
connected by lines that indicate which runs have the same size distribution
of dust particles. The label of run R$xy$ means the particle size
distribution is such that the stopping time ranges $T_{s}\in\lbrack
10^{-x},10^{-y}]$, e.g., in run R30, $T_{s}\in\lbrack10^{-3},10^{0}]$. In each
run, the disk metallicity $Z$ is $0.01$ and $0.03$. To compare the results with our model,
we assume that only the largest particles contribute to turbulence, and that the
effective metallicity is estimated by $Z_{\mathrm{disk}}%
=Z/N_{\mathrm{type}}$, where $N_{\mathrm{type}}=3$ (R10 and R21) or 7 (R30 and
R41) is the number of particle species. We compare our model of $T_{s}=1$ with
runs R10 and R30 (plotted in blue), and the model of
$T_{s}=10^{-1}$ with runs R41 and R21 (plotted in red). $(b)$
Comparison with the radial diffusion coefficient, $D_{x}$, measured in the 3D
runs of the simulation. We assume that the diffusion coefficient of the gas is
represented by that of the smallest particles in the simulation. The values of
$D_{x}$ for the smallest particles are read from Fig.9 of Bai \& Stone (2010).
}%
\label{fig:alpha-bai}%
\end{figure}

Bai \& Stone (2010) performed a three-dimensional simulation, focusing on
investigating turbulence induced by streaming instability. They found that
the streaming instability induced turbulence before the KH instability set in.
They measured the turbulent diffusion coefficient. However, it is difficult to
compare our model directly with the results in Bai \& Stone (2010), because their
simulation includes particles of several sizes (3-7 species) while our model
considers only single-sized particles. In the simulation, it was reported that
only large particles were responsible for inducing turbulence. Our model also
shows that the energy liberation per unit mass of the dust is higher for
larger particles (it is proportional to $T_{s}$ for $T_{s}\la1$). In order to
compare results, we assume that in the simulation, the turbulence is induced
only by the largest particles (i.e., particles of largest $T_{s}$). For
example, in the R41 run (in which the stopping time of the particles ranges
from $10^{-4}$ to $10^{-1}$), we assume that only $T_{s}=0.1$ particles
are responsible for turbulence. We then compare the simulation result with our
model of $T_{s}=0.1$. In the simulation, each species has the same amount of
mass. Because we consider the largest particles only, the total amount of the
dust participating in driving turbulence is $\Sigma
_{d}/N_{\mathrm{type}}$, where $N_{\mathrm{type}}$ is the number of particle
species in the simulation. For example, in the R41 run, $N_{\mathrm{type}}=7$,
and we compare the simulation with a disk metallicity $Z=0.01$ with our
model of $Z_{\mathrm{disk}}=Z/N_{\mathrm{type}}=1.43\times10^{-3}$. 

Figure \ref{fig:alpha-bai}$a$ shows a comparison of the diffusion
coefficient obtained from the simulation ($D_{g,z}(\mathrm{3D})$ in
Table 2 of Bai \& Stone (2010)) with our model (calculated with the
parameter $\tilde{\eta}=0.05^{2}$, which was adopted by Bai \& Stone
(2010)). Although our model of $T_{s}=0.1$ agrees with simulations R41
and R21 (plotted in red), we note a qualitative discrepancy between the $T_{s}=1$ model
and simulations R30 and R10 (plotted in blue). The simulations indicate that the diffusion
coefficient decreases with the disk metallicity, and that its value for
$Z=0.03$ is about an order of magnitude smaller than the value from our 
model. This discrepancy cannot be resolved, even by varying the turnover time
of the largest eddies $T_{e}$ (the dashed line). One possible cause for the
inconsistency is particle clumping and concentration in turbulent eddies,
which are not included in our model. The simulation shows strong
particle clumping when $Z=0.03$ in the R10 run and also temporal clumping for
$Z=0.03$ in the R30 run. Such clumping of particles in turbulent eddies may
suppress diffusion of particles compared to the no-clumping cases of
$Z=0.01$ and could be a cause of a decrease in $\alpha$ when
$Z$ is increased in the simulation. Even if the
largest particles (of $T_{s}\sim1$) concentrate in clumps, the smallest
particles (of $T_{s}\la 0.1$) do not clump, and continue to follow
the turbulent diffusion of the gas (Fig. 7 of Bai \& Stone 2010). In Figure
\ref{fig:alpha-bai}$b$, the turbulent diffusion coefficient in the ``radial
direction'' of the smallest particles in the simulation, $D_{x}$, is compared
with the ``vertical'' diffusion coefficient $\alpha$ in our model. Note that
we compare diffusion coefficients in the different
directions. Since the smallest particles spread out to high altitudes where
turbulence is weak, it is difficult to measure the vertical diffusion
coefficient for the smallest particles in the simulation. In Figure
\ref{fig:alpha-bai}$b$, though we still see a discrepancy compared with
the R30 run, our model appears more consistent with the simulation
results, suggesting that our model properly predicts the ``gas''
diffusion coefficient. 

\section{DISCUSSION}
\label{sec:discuss}

\subsection{The Radial Drift Velocity and Collision Velocity of Dust
Particles}
\label{sec:dustvelocity}

\begin{figure}[ptb]
\epsscale{1.1} \plotone{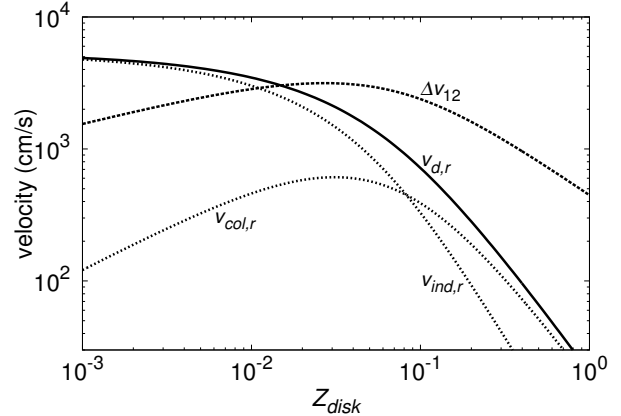} \caption{The radial drift velocity, $\bar
{v}_{d,r}$ (solid line), and the relative velocity of particles due to
turbulence, $\Delta v_{12}$ (dashed line). The contributions of individual
drag, $\bar{v}_{\mathrm{ind},r}$, and collective drag, $\bar{v}_{\mathrm{col}%
,r}$, to the radial drift velocity are plotted with dotted lines.}%
\label{fig:velocity}%
\end{figure}

If the dust-to-gas ratio in the dust layer were larger than unity, the radial
drift velocity of dust particles would be lower than the value that the
particles would have in a gas-rich environment because the gas drag
force could not accelerate sufficiently against the large inertia of the
dust. This effect was pointed out by NSH86 and is seen in equation
(\ref{eq:vdr1}) for the radial velocity due to individual drag. If the
dust-to-gas ratio were much larger than unity, dust accretion would
be caused by collective drag exerted from the slower-orbiting
upper gas layer (Weidenschilling 2003). The radial drift velocity of the
dust is thus a  function of the turbulence strength, $\alpha$. For
weaker turbulence, the 
dust-to-gas ratio in the dust layer is higher, and the individual drag is
weaker. The collective drag is also weaker at smaller $\alpha$ because the
Reynolds stress $P_{\theta z}$ is proportional to $\alpha$. The relative
velocity (or collision velocity) of the dust particles is also a function of
$\alpha$. The radial drift and collision velocities are the important factors in
the dust growth process. In the previous sections, the turbulence strength
$\alpha$ and the dust layer thickness $h_{d}$ have been determined
self-consistently. Using these results, the radial drift velocity and collision
velocity of the dust particles are estimated.

The radial drift velocity is calculated separately for the components due to
individual drag and due to collective drag. For each component, the radial
drift velocity is averaged in the vertical direction. First, the averaged
value of the radial drift velocity due to individual drag is calculated
from equation (\ref{eq:vdr1}),
\begin{eqnarray}
\overline{v}_{\mathrm{ind},r} &=& \frac{1}{\Sigma_{d}}\int_{-\infty}^{\infty}%
\rho_{d}v_{d,r}dz \nonumber \\
&=& -2\eta v_{\mathrm{K}}T_{s}\frac{1}{\Sigma_{d}}\int_{-\infty}^{\infty
}\rho_{d}\frac{1}{T_{s}^{2}+\beta^{2}}dz~.
\end{eqnarray}
The radial drift velocity due to collective drag is calculated from the
vertically-averaged angular momentum loss of the dust component. Integrating
equation (\ref{eq:dlddt}) gives,%
\begin{equation}
\frac{dL_{d}}{dt}=\int_{-\infty}^{\infty}\frac{\partial l_{d}}{\partial
t}dz=-2\eta v_{\mathrm{K}}^{2}T_{s}\Sigma_{d,\mathrm{vis}}~,
\end{equation}
where $\Sigma_{d,\mathrm{vis}}$ is given by equation (\ref{eq:sig_vis2}). This
angular momentum loss causes a radial drift velocity $\bar{v}_{\mathrm{col},r}$ given by
\begin{equation}
\bar{v}_{\mathrm{col},r}=\frac{2}{v_{\mathrm{K}}\Sigma_{d}}\frac{dL_{d}}%
{dt}=-4\eta v_{\mathrm{K}}T_{s}\frac{\Sigma_{d,\mathrm{vis}}}{\Sigma_{d}}~.
\end{equation}
The total radial drift velocity is $\bar{v}_{d,r}=\overline{v}_{\mathrm{ind}%
,r}+\bar{v}_{\mathrm{col},r}$, and is shown in Figure \ref{fig:velocity}
as a solid line, for the case in which the particle size is chosen to maximize the
radial velocity ($T_{s}=1$). In plotting this figure, we adopt
the model parameters at 1 AU of the minimum-mass-solar-nebula model of
Hayashi (1981): $h_{g}/r=0.0326$, $\eta=1.80\times10^{-3}$, and
$\tilde{\eta}=2.92 \times 10^{-3}$. For a small disk metallicity
$Z_{\mathrm{disk}}$, the radial drift velocity is as large 
as 50 m s$^{-1}$, and it decreases with $Z_{\mathrm{disk}}$. For $Z_{\mathrm{disk}}>0.08$, the radial velocity is dominated by collective drag (see the dotted
lines) as pointed out by Weidenschilling (2003). For such large
$Z_{\mathrm{disk}}$, the radial drift velocity due to collective drag also
decreases with $Z_{\mathrm{disk}}$, and then it becomes as small as $1$ m s$^{-1}$
for $Z_{\mathrm{disk}}=0.2$. The radial drift velocity is strongly suppressed
for a sufficiently massive dust layer.

The relative velocity of dust particles due to turbulence, $\Delta v_{12}$, is
calculated for $T_{s}=1$ by substituting $\mathrm{St}_{1}=1$ and
$\mathrm{St}_{2}=0$ into equation (29) in Ormel \& Cuzzi (2007), and is shown
in Figure \ref{fig:velocity} as a dashed line. The collision velocity is
estimated by the larger of $\Delta v_{12}$ and $\bar{v}_{d,r}$. For small
disk metallicities $Z_{\mathrm{disk}}$, the collision velocity is dominated by
the radial drift and is as large as 50 m s$^{-1}$, while for large $Z_{\mathrm{disk}}>0.02$, it is dominated by turbulence. The maximum value of the collision
velocity due to turbulence is about 30 m s$^{-1}$, and it decreases with
$Z_{\mathrm{disk}}$ for $Z_{\mathrm{disk}}>0.03$. Thus, if the dust
particles could survive collisions of $30$ m s$^{-1}$, as suggested by the numerical
simulation of collisions of dust aggregates (Wada et al. 2010) and
$Z_{\mathrm{disk}}\ga0.03$, the dust particles would be able to grow without
being reduced to small fragments.

\subsection{Radial Dependence of the Midplane Dust-to-Gas Ratio}
\label{sec:disketa}

\begin{figure}[ptb]
\epsscale{1.1} \plotone{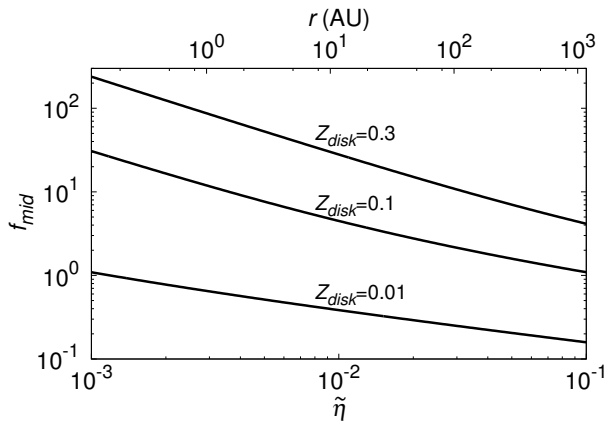} \caption{The midplane dust-to-gas
    ratio $f_{\mathrm{mid}}$ is plotted against $\tilde{\eta}$ for various values
  of the disk metallicity $Z_{\rm disk}$. The top axis shows the
  corresponding radii in the gas disk model by Hayashi (1981), in which
  the disk temperature is $T = 278 (r/{\rm AU})^{-1/2}$K. The dust
  particles are assumed to be small ($T_s \ll 1$) such that
  $f_{\mathrm{mid}}$ is independent of $T_s$.}
\label{fig:fmid-eta}%
\end{figure}

As shown in \S\ref{sec:turbstr}, the turbulence strength and the dust
layer structure depend on properties of the gas disk only through
$\tilde{\eta}$. For larger $\tilde{\eta}$, the accretion velocity of the
dust is faster, and consequently the dust layer is thicker due to
stronger turbulence. Figure \ref{fig:fmid-eta} shows how the midplane
dust-to-gas ratio varies with $\tilde{\eta}$. The figure is plotted in the limit
of $T_s \ll 1$. For such small particles, $f_{\rm mid}$ is
independent of $T_s$ (see Fig. \ref{fig:fdmid}). The midplane
dust-to-gas ratio $f_{\rm mid}$ decreases with $\tilde{\eta}$ as shown
in equation (\ref{eq:fdmid1}). For a gas disk model with a power-law temperature
profile, $T \propto r^{-q}$, $\tilde{\eta}$ behaves as
$\tilde{\eta} \propto r^{1-q}$. The top axis of Figure
\ref{fig:fmid-eta} indicates corresponding locations in the disk
model by Hayashi (1981), i.e.,  $T = 278 (r/{\rm AU})^{-1/2}$K. In a
disk with the standard value of the disk metallicity $Z_{\rm disk}=0.01$,
$f_{\rm mid}$ is less than unity in the most part of the disk except $r
\la 0.1$AU. In disks with $Z_{\rm disk}=0.3$, $f_{\rm mid}$
exceeds $100$ for $r \la 1$AU. We discuss the condition for planetesimal
formation through gravitational instability of the dust layer. In a gas
disk with a surface density profile $\Sigma_g \propto r^{-p}$, the
midplane gas density scales as $\rho_g \propto r^{-p+q/2-3/2}$. The
condition that the dust density exceeds the Roche density ($\rho_{\rm R}
\propto r^{-3}$) is $f_{\rm mid} > \rho_{\rm R}/\rho_g \propto
r^{p-q/2-3/2} \propto r^{p-7/4}$ in a disk model with
$q=1/2$. From the lower line of equation (\ref{eq:fdmid1}), $f_{\rm
  mid}$ decreases as $f_{\rm mid} \propto r^{-(1-q)/(2-\delta)}\propto
r^{-0.47}$. If $p>1.3$, the inner part of the disk is more suitable
for gravitational instability, and vice versa.

\subsection{Does the Liberated Gravitational Energy Go into
Turbulence?}
\label{sec:ene-into-tur}

In this paper, we calculate the liberated gravitational energy from dust
accretion, assuming some fraction of the liberated energy is transferred to
turbulence. The estimate of the dust accretion rate is based on the formula for the
particle terminal velocity derived in NSH86. ``The terminal velocity'' means
that all the liberated energy is consumed by gas drag, converting directly into
the thermal energy of the dust particles and of the surrounding gas molecules.
Thus, one may expect that only a small fraction (or nothing) of the liberated
energy would be used for maintaining turbulence. However, a comparison with
the simulation of KH instability by JHK06 shows that the efficiency factor
$C_{\mathrm{eff}}\approx0.2$ is not negligibly small. In the following subsection, we
discuss the validity of using the particle terminal velocity for calculating
the energy supply rate to turbulence. The energy liberation rate from the
accreting dust calculated in \S \ref{sec:eneinp} is compared with the deposit
rate of the free energy that is the source of several instabilities, such as KH
instability and streaming instability.

\subsubsection{Kelvin-Helmholtz Instability}

The free energy that induces the KH instability originates from the velocity
difference between the midplane dust layer and the upper gas layer, and is
stored as the dust particles settle to the midplane. We estimate the deposit
rate of the free energy during dust sedimentation, and show that it has the
same order of magnitude as the energy liberation rate from the dust accretion
towards the star. Consider two states of dust distribution: the
initial state, in which the dust particles are distributed uniformly in the gas
disk, and the final state, in which all the dust has settled at
the midplane. In the initial state, there is no vertical shear in the disk, and in the final state, the velocity difference $\Delta
v_{\theta}=\eta v_{\mathrm{K}}$ appears between the midplane dust layer and
the upper gas layer. The free energy for KH instability is $\Delta
E_{\mathrm{KH}}\sim\frac{1}{2}\Sigma_{d}\Delta v^{2}\sim\frac{1}{2}\Sigma
_{d}\eta^{2}v_{\mathrm{K}}{}^{2}$, for $\Sigma_{d}\ll\Sigma_{g}$. The settling
timescale is $\tau_{\mathrm{sed}}=(T_{s}^{2}+1)/(T_{s} \Omega_{\mathrm{K}})$,
and then the energy deposit rate is
\begin{equation}
\frac{\Delta E_{\mathrm{KH}}}{\tau_{\mathrm{sed}}}\sim\frac{1}{2}\eta
^{2}v_{\mathrm{K}}^{2} \Omega_{\mathrm{K}} \frac{T_{s}}{T_{s}^{2}+1} \Sigma_{d} \ ,
\end{equation}
which is the same order as the energy liberation rate of the accreting dust
(eqs. [\ref{eq:ene_drag}] and [\ref{eq:sig_drag2}]). Hence, the deposition rate
of the free energy for KH instability can be estimated by the energy
liberation rate of the accreting dust.

\subsubsection{Streaming Instability}
\label{sec:streaminst}

The free energy for streaming instability originates from the velocity
difference between the dust particles and the surrounding gas. When streaming
instability begins, the velocity difference decreases as the free energy is
consumed by inducing turbulence. In fact, this decrease in the velocity difference
can be seen even in the linear growth regime. Youdin \& Goodman (2005)
showed in their Figure 6 that the velocity difference between the dust and gas
decreases (increases) at the locations where the particle density increases
(decreases). The spatially averaged value of the free energy decreases as the
perturbation grows. Thus, without an energy supply, streaming instability would
cease. Given a state in which the velocity difference between the dust and the
gas has reduced, the dust particles are no longer in equilibrium: the
gravity, the centrifugal force, and the gas drag force are not in balance. The
dust particles are accelerated and the velocity difference from the gas rises
again. To estimate the effect of the energy deposition on the velocity
difference, we consider a state in which the dust density is similar to the
gas density, $\rho_{d}\sim\rho_{g}$. In such a state, streaming
instability occurs efficiently with a growth time of the order of the
Keplerian time (for the short wave branch, Youdin \& Goodman 2005; Youdin \&
Johansen 2007). For dust particles of $T_{s}<1$, the terminal velocities of
the dust and of the gas are of the order of $T_{s}\eta v_{\mathrm{K}}$ (eq.
[\ref{eq:vdr}], [\ref{eq:vgr}]), and thus, the free energy per unit area is
$\Delta E_{\mathrm{str}}\sim\Sigma_{d}T_{s}^{2}\eta^{2}v_{\mathrm{K}}^{2}$. This
deposition of free energy occurs during the acceleration phase of the dust,
and thus in the stopping time $T_{s}\Omega_{\mathrm{K}}^{-1}$, and then the energy is
transferred to turbulence in the growth timescale $\Omega_{\mathrm{K}}^{-1}$ of
streaming instability. Thus, the timescale for the energy deposit in
turbulence is the sum of these timescales, and for $T_{s}\la1$, $\tau
_{\mathrm{str}}=T_{s}\Omega_{\mathrm{K}}^{-1}+\Omega_{\mathrm{K}}^{-1}\sim\Omega_{\mathrm{K}}^{-1}$. The
energy deposition rate is
\begin{equation}
\frac{\Delta E_{\mathrm{str}}}{\tau_{\mathrm{str}}}\sim\eta^{2}
v_{\mathrm{K}}^{2} \Omega_{\mathrm{K}} T_{s}^{2}\Sigma_{d} \ ,
\end{equation}
which is smaller by a factor $T_{s}$ than the estimate from the dust accretion
rate (eq.[\ref{eq:ene_drag}]). Hence, our estimate, derived from the dust accretion
rate, is appropriate for particles of $T_{s}\sim1$. For smaller particles
($T_{s} \ll1$), the energy deposition rate is higher for KH instability than for
streaming instability, and KH instability is expected to operate first. The
energy deposition rate for instability (either for KH or streaming instabilities)
is estimated from the energy liberation rate due to dust accretion.

\section{SUMMARY}

In this paper, we discuss turbulence induced in the dust layer. The
 turbulence strength or the parameter $\alpha$ is determined using the energetics of
dust accretion towards the central star. The key concept is that the dust
particles reside in a deeper potential than the gas. The effective potential,
including the gas pressure, is $-GM(1-2\eta)/r$ for the gas,
and $-GM/r$ for the dust. When angular momentum is transferred
from the dust to the gas through gas drag, the dust particles lose more energy
than the gas gains. The excess energy can be used for exciting
turbulence. If the dust accretion due to gas drag is a primary source of energy
liberation, i.e., if the gas accretion rate due to turbulence is much smaller
than the dust accretion rate, then the turbulence strength is determined by the
energy supply rate from the dust accretion. This is not the case if the gas
disk itself is turbulent via, e.g., MRI. If the dust layer is composed of large
particles with stopping time $T_{s}\gg1$, then the gas accretion may
dominate the dust accretion, as discussed in \S \ref{sec:coldrag}.

 We estimate the dust accretion rate using the terminal velocity profiles of the
dust particles in a laminar disk derived by NSH86. The expected turbulence
strength and corresponding structure of the dust layer from our analysis agree
with the previous analytical result on the marginally KH-unstable dust layer
by S98. As our analysis does not assume tight coupling of the dust to the
gas, nor specify the mechanism of instability that induces turbulence, it is
considered an extension of the analysis of S98 to a more general physical
situation of the dust layer. The results of this paper agree with
the results in Michikoshi \& Inutsuka (2006), which analyzes KH instability
of the dust layer composed of particles with large stopping times ($T_{s}>1$),
as shown in Figure \ref{fig:michi}. 

Our analysis shows that, for particles of
$T_{s}\la1$, the turbulence strength is smaller than $\alpha_{\max}\sim
C_{\mathrm{eff}}{\tilde{\eta}}T_{s}$, where $C_{\mathrm{eff}}=0.19$ is the
efficiency of the energy supply to turbulence (see Fig. \ref{fig:alpha} and
eq. [\ref{eq:alpha2}]). This strength reaches a maximum when the disk
metallicity is $Z_{\mathrm{disk}}\sim\sqrt{C_{\mathrm{eff}}\tilde{\eta
}}\sim10^{-2}$. Modifying the disk metallicity from the standard value,
$10^{-2}$, by any process, results in weaker turbulence and a thinner dust
layer, and consequently may accelerate the growth process of the dust
particles, as pointed out in S98.

Comparison of our results with previous numerical simulations of KH and streaming
instabilities by JHK06 and Bai \& Stone (2010) shows quantitative agreement
with our analysis for dust particles of $T_{s}\la0.1$, although there may be a
qualitative disagreement for $T_{s}\ga1$ particles (see Figs.
\ref{fig:alpha-joh} $-$ \ref{fig:alpha-bai}). Hence, we conclude that
turbulence in the dust layer is controlled by the energy supply from the dust
accretion due to gas drag, provided that the dust particles are not so large
that $T_{s}\ga1$. In such a layer, turbulence strength is estimated by the dust
accretion rate (eq.[\ref{eq:alpha}]).

\acknowledgements
This work was stimulated by discussions with Minoru Sekiya. We are
grateful to Anders Johansen for providing detailed information on his
simulations, and to Chris Ormel and Eugene Chiang for useful
discussions. We also thank an anonymous referee for helpful comments.
This work was supported in part by  Grants-in-Aid for Scientific Research,
Nos. 20540232, 22$\cdot$2942, and 22$\cdot$7006 from the Ministry of
Education, Culture, Sports, Science, and Technology (MEXT), Japan.

\appendix

\section{DUST AND GAS VELOCITIES IN STEADY LAMINAR FLOW}
\label{sec:app-vel}

We present calculations of the dust and gas velocities in steady laminar flow in this
appendix. We follow NSH86, but extend their calculation to the second order
of $\eta$. The equations of motion of the gas and of the dust are,
respectively,%
\begin{equation}
\frac{d\mbox{\boldmath$v$}_{g}}{dt}=-\frac{GM}{r^{3}}\mbox{\boldmath$r$}-\frac{1}%
{\rho_{g}}\nabla P-\frac{\rho_{d}}{\rho_{g}}\frac{\Omega_{\mathrm{K}}}{T_{s}%
}(\mbox{\boldmath$v$}_{g}-\mbox{\boldmath$v$}_{d})~, 
\label{eq:motion-g-v}%
\end{equation}%
\begin{equation}
\frac{d\mbox{\boldmath$v$}_{d}}{dt}=-\frac{GM}{r^{3}}\mbox{\boldmath$r$}-\frac
{\Omega_{\mathrm{K}}}{T_{s}}(\mbox{\boldmath$v$}_{d}-\mbox{\boldmath$v$}_{g})~,
\label{eq:motion-d-v}%
\end{equation}
where $\mbox{\boldmath$v$}_{g}$ and $\mbox{\boldmath$v$}_{d}$ are the velocities of the
gas and the dust. The radial and azimuthal components of the velocity in the
cylindrical coordinates $(r,\theta)$ are normalized by the Keplerian velocity
$v_{\mathrm{K}}$, such as $v_{g,r}=\tilde{v}_{g,r}v_{\mathrm{K}}$,
$v_{g,\theta}=\tilde{v}_{g,\theta}v_{\mathrm{K}}$. We assume that the
velocities, $v_{g,r}$ etc., vary with $r$ in the same way as the Keplerian
velocity $v_{\mathrm{K}}$, i.e., that the non-dimensional velocities, $\tilde
{v}_{g,r}$ etc., are constant with $r$. This assumption is satisfied when
$\rho_{d}/\rho_{g}$ and $\eta$ are constant with $r$.(see eqs.[\ref{eq:vgr1}%
]-[\ref{eq:vdt2}] below) Then, the radial derivative of the velocity is, for
example,%
\begin{equation}
\frac{\partial}{\partial r}v_{g,r}=-\frac{v_{g,r}}{2r}~,
\end{equation}
and the radial derivative of other velocity components has a similar
form. In a steady axisymmetric state ($\partial/\partial t=\partial
/\partial\theta=0$), equations (\ref{eq:motion-g-v}) and (\ref{eq:motion-d-v})
become%
\begin{equation}
-\frac{1}{2}\tilde{v}_{g,r}^{2}-\tilde{v}_{g,\theta}^{2}=-(1-2\eta)-\frac
{\rho_{d}}{\rho_{g}}\frac{1}{T_{s}}(\tilde{v}_{g,r}-\tilde{v}_{d,r})~,
\label{eq:motion-g-r}%
\end{equation}%
\begin{equation}
\frac{1}{2}\tilde{v}_{g,r}\tilde{v}_{g,\theta}=-\frac{\rho_{d}}{\rho_{g}}%
\frac{1}{T_{s}}(\tilde{v}_{g,\theta}-\tilde{v}_{d,\theta})~,
\label{eq:motion-g-t}%
\end{equation}%
\begin{equation}
-\frac{1}{2}\tilde{v}_{d,r}^{2}-\tilde{v}_{d,\theta}^{2}=-1-\frac{1}{T_{s}%
}(\tilde{v}_{d,r}-\tilde{v}_{g,r})~, 
\label{eq:motion-d-r}%
\end{equation}%
\begin{equation}
\frac{1}{2}\tilde{v}_{d,r}\tilde{v}_{d,\theta}=-\frac{1}{T_{s}}(\tilde
{v}_{d,\theta}-\tilde{v}_{g,\theta})~. 
\label{eq:motion-d-t}%
\end{equation}
The non-dimensional velocities are expanded in a power series of $\eta$,%
\begin{equation}
\tilde{v}_{g,r}=\tilde{v}_{g,r,1}\eta+\tilde{v}_{g,r,2}\eta^{2}+O(\eta^{3})~,
\end{equation}%
\begin{equation}
\tilde{v}_{g,\theta}=1+\tilde{v}_{g,\theta,1}\eta+\tilde{v}_{g,\theta,2}%
\eta^{2}+O(\eta^{3})~,
\end{equation}%
\begin{equation}
\tilde{v}_{d,r}=\tilde{v}_{d,r,1}\eta+\tilde{v}_{d,r,2}\eta^{2}+O(\eta^{3})~,
\end{equation}%
\begin{equation}
\tilde{v}_{d,\theta}=1+\tilde{v}_{d,\theta,1}\eta+\tilde{v}_{d,\theta,2}%
\eta^{2}+O(\eta^{3})~.
\end{equation}
Substituting the above expressions into equations (\ref{eq:motion-g-r}%
)-(\ref{eq:motion-d-t}) yields in the first order of $\eta,$%
\begin{equation}
\tilde{v}_{g,r,1}=\frac{\rho_{d}}{\rho_{g}}\frac{2T_{s}}{T_{s}^{2}+\beta^{2}%
}~, 
\label{eq:vgr1}%
\end{equation}%
\begin{equation}
\tilde{v}_{g,\theta,1}=-\frac{T_{s}^{2}+\beta}{T_{s}^{2}+\beta^{2}}~,
\end{equation}%
\begin{equation}
\tilde{v}_{d,r,1}=-\frac{2T_{s}}{T_{s}^{2}+\beta^{2}}~, 
\label{eq:vdr1}%
\end{equation}%
\begin{equation}
\tilde{v}_{d,\theta,1}=-\frac{\beta}{T_{s}^{2}+\beta^{2}}~,
\end{equation}
which are the same as the results of NSH86. In the second order of $\eta$,%
\begin{equation}
\tilde{v}_{g,r,2}=\frac{\rho_{d}}{\rho_{g}}\frac{T_{s}^{3}{}(3T_{s}^{2}%
{}+2\beta^{2}+2\beta)}{\left(  T_{s}^{2}{}+\beta^{2}\right)  ^{3}}~,
\end{equation}%
\begin{equation}
\tilde{v}_{g,\theta,2}=-\frac{T_{s}^{6}{}+3\beta T_{s}^{4}{}+3\beta^{2}%
T_{s}^{2}{}+\beta^{4}}{2\left(  T_{s}^{2}{}+\beta^{2}\right)  ^{3}}~,
\end{equation}%
\begin{equation}
\tilde{v}_{d,r,2}=-\frac{T_{s}^{3}{}(T_{s}^{2}{}+2\beta)}{\left(  T_{s}^{2}%
{}+\beta^{2}\right)  ^{3}}~,
\end{equation}%
\begin{equation}
\tilde{v}_{d,\theta,2}=-\frac{\beta\left(  3T_{s}^{4}{}+2\beta^{2}T_{s}^{2}%
{}+3\beta T_{s}^{2}{}+\beta^{3}\right)  }{2\left(  T_{s}^{2}{}+\beta
^{2}\right)  ^{3}}~~. 
\label{eq:vdt2}%
\end{equation}

\section{ENERGY LIBERATION RATE DUE TO INDIVIDUAL DRAG}
\label{sec:app-ind}

In this appendix, we describe a more rigorous derivation of the energy
liberation rate due to individual drag than was provided in
\S \ref{sec:inddrag}. In a laminar disk, the particle drift velocity $v_{d,r}$
calculations were presented in NSH86, to the first order of
$\eta$. Since the liberated energy is of the second order of $\eta$ (see eq.
[\ref{eq:ene_drag2}] below), we use the particle radial velocity, which is
calculated to the order of $\eta^{2}$ in Appendix \ref{sec:app-vel},
\begin{equation}
v_{d,r}=-\left[  \frac{2{T_{s}}}{{T_{s}^{2}+\beta^{2}}}\eta+\frac{T_{s}%
^{3}\left(  T_{s}^{2}+2\beta\right)  }{\left(  T_{s}^{2}+\beta^{2}\right)
^{3}}\eta^{2}\right]  v_{\mathrm{K}}\ , 
\label{eq:vdr2}%
\end{equation}
where the first term corresponds to equation (2.11) in NSH86. The gas drifts in
the opposite direction with the velocity $v_{g,r}$,
\begin{equation}
v_{g,r}=\frac{{\rho_{d}}}{{\rho_{g}}}\left[  \frac{2{T_{s}}}{{T_{s}^{2}%
+\beta^{2}}}\eta+\frac{T_{s}^{3}\left(  3T_{s}^{2}+2\beta^{2}+2\beta\right)
}{\left(  T_{s}^{2}+\beta^{2}\right)  ^{3}}\eta^{2}\right]  v_{\mathrm{K}}\ .
\label{eq:vgr2}%
\end{equation}
The liberated gravitational energy per unit surface area of the disk is,%
\begin{equation}
\frac{\partial E_{\mathrm{drag}}}{\partial t}=\frac{1}{2}\int_{-\infty
}^{\infty}(\rho_{d}{g_{d}v_{d,r}+}\rho_{g}{g_{g}v_{g,r})dz}~,
\end{equation}
where the factor of $1/2$ accounts for the work used for the acceleration (and
deceleration) of the azimuthal velocity of the dust (and of the gas) as their
semi-major axes change. Using equation (\ref{eq:vdr2}) and (\ref{eq:vgr2}),
the energy liberation rate is given by
\begin{equation}
\frac{\partial E_{\mathrm{drag}}}{\partial t}=2\eta^{2}v_{\mathrm{K}}%
^{2}\Omega_{\mathrm{K}}T_{s}\Sigma_{d,\mathrm{drag}}\ , 
\label{eq:ene_drag2}%
\end{equation}
where the effective ``surface density'' of the dust is
\begin{equation}
\Sigma_{d,\mathrm{drag}}=\frac{{\Sigma_{d}}}{\sqrt{\pi}}\int_{-\infty}%
^{\infty}{\frac{{\exp(-\tilde{z}^{2})}}{{T_{s}^{2}+\beta}^{2}}\left[
1-\frac{T_{s}^{2}}{2(T_{s}^{2}+{\beta}^{2})}\right]  d\tilde{z}}\ .
\label{eq:sig_drag_ap}%
\end{equation}

\section{ENERGY LIBERATION RATE DUE TO COLLECTIVE DRAG}
\label{sec:app-col}

In this appendix, we provide a more rigorous derivation of the energy
liberation rate due to collective drag than was given in \S
\ref{sec:coldrag}. The orbital velocity of the gas presented in NSH86 is 
\begin{equation}
v_{g,\theta}=\left(  1-\frac{{\beta+T_{s}^{2}}}{{\beta^{2}+T_{s}^{2}}}%
\eta\right)  v_{\mathrm{K}}\ ,
\end{equation}
where $\beta=(\rho_{d}+\rho_{g})/\rho_{g}$ varies with the altitude $z$. The
$\theta z$ component of the Reynolds stress $P_{\theta z}$ due to the
turbulent viscosity $\nu$ of the gas is
\begin{equation}
P_{\theta z}=(\rho_{g}+C_{\mathrm{str}}\rho_{d})\nu\frac{{\partial
v_{g,\theta}}}{{\partial z}}\ .
\end{equation}
We add the factor $C_{\mathrm{str}}$ to account for the weaker coupling of
the dust to the gas for larger dust particles. YL07 has shown that the
contribution of the dust to the $r\theta$-component of the Reynolds stress,
$P_{r\theta}$, is a factor $1/(T_{s}^{2}+1)$ times the gas contribution. From
equations (33c) and (B.1) of YL07, it is seen that in both limits of
$T_{e}\ll1$ and $T_{e}\gg1$, $\left\langle v_{d,r}^{\prime}v_{d,\theta
}^{\prime}\right\rangle \sim\left\langle v_{g,r}^{\prime}v_{g,\theta}^{\prime
}\right\rangle T_{s}^{-2}$ for $T_{s}\gg1$, and $\left\langle v_{d,r}^{\prime
}v_{d,\theta}^{\prime}\right\rangle \sim\left\langle v_{g,r}^{\prime
}v_{g,\theta}^{\prime}\right\rangle $ for $T_{s}\ll1$, where the prime denotes
velocity fluctuations. We assume that a similar relationship holds for the $\theta
z$-component of the Reynolds stress, $P_{\theta z}$, and thus $C_{\mathrm{str}%
}$ is expressed as\footnote{Equation (B.1) of YL07 is based on the radial
shear effect, and its applicability for the $\theta z$-component is not very
clear. We simply assume that $P_{r\theta}$ and $P_{\theta z}$ have similar
properties. Note also that $C_{\mathrm{str}}$ includes only the
effect of $T_{s}$, assuming that the dust particles act as passive particles
in the gas turbulence. This is not the case if the local dust-to-gas ratio is
larger than unity. Since the effect of the inertia of the dust
on the turbulence is unclear, we simply adopt equation (\ref{eq:Cstr}). For
$f_{\mathrm{mid}}\ge1$, the simple plate drag approximation (eq.
[\ref{eq:sig_vis-approx}]) may provide a more accurate estimate. The
$\Sigma_{d,\mathrm{vis}}$ estimated from the plate drag approximation (eq.
[\ref{eq:sig_vis-approx}]) and from the calculation in this Appendix (eq.
[\ref{eq:sig_vis2}]) does not suggest a big difference at large $f_{\mathrm{mid}}$. In the plate drag approximation, $\Sigma_{d,\mathrm{vis}}\propto
f_{\mathrm{mid}}^{-1}$, while the calculation in this Appendix gives
$\Sigma_{d,\mathrm{vis}}\propto f_{\mathrm{mid}}^{-0.9}$ (see Fig.
\ref{fig:Sigma}).}
\begin{equation}
C_{\mathrm{str}}=\frac{1}{T_{s}^{2}+1}\ , 
\label{eq:Cstr}%
\end{equation}
This stress transfers angular momentum in the $z$-direction, and the time
derivatives of the angular momentum of the dust and of the gas per unit volume
and unit time are, respectively,
\begin{equation}
\frac{\partial l_{d}}{\partial t}=\frac{{C_{\mathrm{str}}\rho_{d}}}{{\rho
_{g}+C_{\mathrm{str}}\rho_{d}}}r\frac{{\partial P_{\theta z}}}{{\partial z}%
}\ , 
\label{eq:dlddt}%
\end{equation}
and
\begin{equation}
\frac{\partial l_{g}}{\partial t}=\frac{{\rho_{g}}}{{\rho_{g}+C_{\mathrm{str}%
}\rho_{d}}}r\frac{{\partial P_{\theta z}}}{{\partial z}}\ .
\end{equation}
Here, we assume that the viscous torque is distributed to the dust and to the
gas with the ratio $C_{\mathrm{str}}\rho_{d}:\rho_{g}$. The corresponding
energy change is $\partial\varepsilon_{d}/\partial t=\Omega_{\mathrm{K}%
}\partial l_{d}/\partial t$ for the dust and $\partial\varepsilon_{g}/\partial
t=\Omega_{g}\partial l_{g}/\partial t$ for the gas 
\footnote{We assume that the
rotational velocities of the dust and the gas are $\Omega_{\mathrm{K}}$ and
$\Omega_{g}$ respectively, neglecting the modification of the rotational
velocity due to gas drag. This assumption is justified if $f_{\mathrm{mid}}\ge1$ and if the angular momentum exchange occurs between the dust-dominant
layer (rotating with $\Omega_{\mathrm{K}}$) and the gas-dominant layer
(rotating with $\Omega_{g}$). For $f_{\mathrm{mid}}\le1$, the energy
liberation due to collective drag is neglected compared to that due to
individual drag (see Fig. \ref{fig:Sigma}).}.
In sum, the energy liberation rate per unit area and unit time is
\begin{eqnarray}
\frac{\partial E_{\mathrm{vis}}}{\partial t}  & = & -\int_{-\infty}^{\infty}\left(
\frac{\partial\varepsilon_{d}}{\partial t}+\frac{\partial\varepsilon_{g}%
}{\partial t}\right) dz \nonumber \\
& = & -v_{\mathrm{K}}\eta\int_{-\infty}^{\infty}{P_{\theta z}\frac{\partial}{{\partial
z}}\left(  {\frac{1}{{1+C_{str}(\rho_{d}/\rho_{g})}}}\right)  dz} \ ,
\label{eq:ene_vis0}%
\end{eqnarray}
where we use $P_{\theta z}(\pm\infty)=0$, and the minus sign is added to give
the energy release rate. The integration variable is transferred to $\tilde
{z}=z/(\sqrt{2}h_{d})$, and then using
\begin{equation}
\frac{{\partial v_{g,\theta}}}{{\partial\tilde{z}}}=-2f_{\mathrm{mid}}%
\tilde{z}\exp(-\tilde{z}^{2})\frac{{\beta^{2}+2\beta T_{s}^{2}-T_{s}^{2}}%
}{{(T_{s}^{2}+\beta^{2})^{2}}}\eta v_{\mathrm{K}}\ ,
\end{equation}
and
\begin{equation}
\frac{\partial}{{\partial\tilde{z}}}\left(  {\frac{1}{{1+C_{\mathrm{str}}%
(\rho_{d}/\rho_{g})}}}\right)  =\frac{{2C_{\mathrm{str}}f_{\mathrm{mid}}\tilde{z}\exp(-\tilde{z}^{2})}}{{[1+C_{\mathrm{str}}f_{\mathrm{mid}}%
\exp(-\tilde{z}^{2})]^{2}}}\ ,
\end{equation}
equation (\ref{eq:ene_vis0}) becomes
\begin{equation}
\frac{\partial E_{\mathrm{vis}}}{\partial t}=2\eta^{2}v_{\mathrm{K}}^{2}%
\Omega_{\mathrm{K}}T_{s}\Sigma_{d,\mathrm{vis}}\ , 
\label{eq:ene_vis2}%
\end{equation}
where
\begin{equation}
\Sigma_{d,\mathrm{vis}}=\frac{C_{\mathrm{str}}}{\sqrt{{\pi}}}f_{\mathrm{mid}}\Sigma_{d}\frac{{1+2T}_{s}}{1+{T_{s}}}\int_{-\infty}^{\infty}{\frac
{{\beta^{2}+2\beta T_{s}^{2}-T_{s}^{2}}}{{(T_{s}^{2}+\beta^{2})^{2}}}%
\frac{{\tilde{z}^{2}\exp(-2\tilde{z}^{2})}}{{[1+C_{\mathrm{str}}%
f_{\mathrm{mid}}\exp(-\tilde{z}^{2})]}}d\tilde{z}}\ . 
\label{eq:sig_vis2}%
\end{equation}

\section{ENERGY DISSIPATION RATE OF TURBULENT DUST MOTION}
\label{sec:app-ene-dis}

The energy dissipation rate of the gas in turbulence is
\begin{equation}
\varepsilon_{g}=\frac{{u_{g,\mathrm{eddy}}^{2}}}{\tau{_{g,\mathrm{eddy}}}}~,
\label{eq:enedis-gas}%
\end{equation}
where $u_{g,\mathrm{eddy}}$ and $\tau_{g,\mathrm{eddy}}$ are, respectively, the
velocity and the turnover time of the largest eddies. Similarly, for the
dust,
\begin{equation}
\varepsilon_{d}=\frac{{u_{d,\mathrm{eddy}}^{2}}}{\tau{_{d,\mathrm{eddy}}}}~.
\end{equation}
The eddy velocity of the dust, $u_{d,\mathrm{eddy}}$, is estimated from equation
(20) of YL07 as%
\begin{equation}
{u_{d,\mathrm{eddy}}^{2}=}\frac{{u_{g,\mathrm{eddy}}^{2}}}{1+T_{s}T_{e}%
^{-1}+T_{s}T_{e}}~,
\end{equation}
where $T_{e}=\tau_{g,\mathrm{eddy}}\Omega_{\mathrm{K}}$ is the non-dimensional
turnover time of the gas turbulence. The turnover time of the dust turbulence,
$\tau_{d,\mathrm{eddy}}$, is estimated as the larger value of $\tau
_{g,\mathrm{eddy}}$ and $\tau_{\mathrm{stop}}$, i.e.,
\begin{equation}
\tau{_{d,\mathrm{eddy}}=\max(\tau_{g,\mathrm{eddy}},\tau_{\mathrm{stop}})~.}
\label{eq:teddy-dust}%
\end{equation}
From equations (\ref{eq:enedis-gas})-(\ref{eq:teddy-dust}),
\begin{equation}
\varepsilon_{d}=C_{\mathrm{ene}}\varepsilon_{g}~,
\end{equation}
where $C_{\mathrm{ene}}$ is approximately
\begin{equation}
C_{\mathrm{ene}}=\left\{
\begin{array}
[c]{ccc}%
\frac{1}{T_{s}T_{e}+1} & \mathrm{for} & T_{s}\,\leq T_{e}\\
\frac{1}{T_{s}^{2}(T_{e}^{-2}+1)} & \mathrm{for} & T_{s}\,>T_{e}%
\end{array}
\right.  \ .
\end{equation}


\end{document}